\documentstyle[psfig]{mn}

\newif\ifAMStwofonts
\AMStwofontstrue

\def\spose#1{\hbox to 0pt{#1\hss}}
\def \da{\Delta\alpha/\alpha}
\def \hiz{high-$z$}
\def \loz{low-$z$}

\title[Possible evidence for a variable $\alpha$ from QSO absorption lines:
systematic errors] {Possible evidence for a variable fine structure
constant from QSO absorption lines: systematic errors\thanks{Data
presented herein were obtained at the W.M. Keck Observatory, which is
operated as a scientific partnership among the California Institute of
Technology, the University of California and the National Aeronautics and
Space Administration. The Observatory was made possible by the generous
financial support of the W.M. Keck Foundation.}}

\author[M. T. Murphy et al.]  {M. T. Murphy$^1$\thanks{E-mail:
       mim@phys.unsw.edu.au (MTM)}, J. K. Webb$^1$, V. V. Flambaum$^1$,
       C. W. Churchill$^2$ and\newauthor J. X. Prochaska$^3$\\
$^1$School of Physics, The
       University of New South Wales, UNSW Sydney NSW 2052, Australia\\
$^2$Department of Astronomy \& Astrophysics, Pennsylvania State
       University, University Park, PA, 16802, USA\\
$^3$The Observatories
       of the Carnegie Institute of Washington, 813 Santa Barbara
       St. Pasadena, CA 91101}

\date{Accepted ---.
      Received ---;
      in original form ---}

\pagerange{\pageref{firstpage}--\pageref{lastpage}}
\pubyear{2000}

\begin{document}

\maketitle

\label{firstpage}

\begin{abstract}
Comparison of quasar absorption spectra with laboratory spectra allow us to
probe possible variations in the fundamental constants over cosmological
time-scales. In a companion paper we present an analysis of Keck/HIRES
spectra and report possible evidence suggesting that the fine structure
constant, $\alpha$, may have been smaller in the past: $\da = (-0.72 \pm
0.18) \times 10^{-5}$ over the redshift range $0.5 < z < 3.5$.  In this
paper we describe a comprehensive investigation into possible systematic
effects.  Most of these do not significantly influence our results.  When
we correct for those which do produce a significant systematic effect in
the data, the deviation of $\da$ from zero becomes {\it more} significant.
We are lead increasingly to the interpretation that $\alpha$ was slightly
smaller in the past.
\end{abstract}

\begin{keywords}
atomic data -- line: profiles -- instrumentation: spectrographs -- methods:
data analysis -- techniques: spectroscopic -- quasars: absorption lines
\end{keywords}

\section{Introduction}
The idea of varying fundamental constants is not a new one: Milne (1935,
1937) and Dirac (1937) independently suggested a varying Newton
gravitational constant. More modern theories, such as superstring theory,
naturally predict variations in fundamental coupling constants
(e.g. Forg\'{a}cs \& Horv\'{a}th 1979; Marciano 1984; Barrow 1987; Li \&
Gott 1998), strongly motivating an experimental search. Several authors
have demonstrated the idea (Varshalovich \& Potekhin 1995 and references
therein), first used by Bahcall, Sargent \& Schmidt (1967), of using
spectra of high redshift gas clouds seen in absorption against background
quasars (QSOs) to constrain possible variation of the fine structure
constant, $\alpha\equiv\frac{e^2}{\hbar c}$, over cosmological
time-scales. This method is based on a comparison of the observed
transition wavelengths of alkali doublets and recent applications of it
have yielded constraints on $\da\equiv(\alpha_z-\alpha_0)/\alpha_0 \sim 5
\times 10^{-5}$ where $\alpha_z$ and $\alpha_0$ are the values of $\alpha$
at the absorption cloud redshift $z$ and in the laboratory respectively
(e.g. Cowie \& Songaila 1995; Ivanchik, Potekhin \& Varshalovich 1999;
Varshalovich, Potekhin \& Ivanchik 2000). Recently, we have increased the
level of precision obtained with the alkali doublet method to $\da = (-0.5
\pm 1.3) \times 10^{-5}$ using higher quality data (Murphy et al. 2000c).

However, a new method offering an order of magnitude improvement in
precision was suggested by Dzuba, Flambaum \& Webb (1999a,b) and was
demonstrated by Webb et al. (1999, hereafter W99). The increase in
precision derives from the fact that the relativistic corrections to the
energy levels of different ions vary from species to species by up to an
order of magnitude. A comparison of the observed Mg{\sc \,i}, Mg{\sc \,ii}
and Fe{\sc \,ii} transition wavelengths in 30 absorption systems allowed
W99 to tentatively suggest that the mean value of $\da = (-1.09 \pm
0.36)\times 10^{-5}$ over the redshift range $0.5 < z < 1.6$: the fine
structure constant may have been smaller in the past.

The present work is the companion paper of Murphy et al. (2001a, hereafter
M01a) in which we discuss our techniques in detail and present the results
of recent work. We summarise the main points here.  We obtained Keck/HIRES
spectra of 28 QSOs and analysed 49 absorption systems lying along their
lines of sight. These systems included those Mg{\sc \,ii}/Fe{\sc \,ii}
systems considered in W99 and also a higher redshift sample of damped
Lyman-$\alpha$ systems (DLAs) containing absorption lines of ions such as
Mg{\sc \,i}, Mg{\sc \,ii}, Al{\sc \,ii}, Al{\sc \,iii}, Si{\sc \,ii},
Cr{\sc \,ii}, Fe{\sc \,ii}, Ni{\sc \,ii} and Zn{\sc \,ii}, so that our
redshift range now covers $0.5<z<3.5$, corresponding to look-back times
from 5 to 12 billion years ($H_0=68{\rm ~kms}^{-1}{\rm Mpc}^{-1}$,
$\Omega_{\rm M}=0.3$, $\Omega_{\Lambda}=0.7)$. Our absorption systems fell
conveniently into two sub-samples: the Mg{\sc \,ii}/Fe{\sc \,ii} systems
comprised the low redshift sample ($\bar{z}=1.0$) and the DLAs comprised
the high redshift sample ($\bar{z}=2.1$). There were, however, 2 Mg{\sc
\,ii}/Fe{\sc \,ii} absorbers within the \hiz~sample that lay at low
redshift but we have included them in the \hiz~sample since the
observations and data reduction for the two sub-samples were carried out
independently by two different groups.

The analysis is based on the following equation for the wavenumber,
$\omega_z$, of a given transition at a redshift $z$;
\begin{equation}
\label{eq:omega}
\omega_z = \omega_0 + q_1x + q_2y\, ,
\end{equation}
where $\omega_0$ is the laboratory rest wavenumber of the transition, $q_1$
and $q_2$ are relativistic coefficients, $x\equiv(\alpha_z/\alpha_0)^2-1$
and $y \equiv (\alpha_z/\alpha_0)^4-1$. Note that the second and third
terms on the right hand side only contribute if $\da\neq 0$. The values of
$q_1$ and $q_2$ have been calculated using many-body techniques in Dzuba et
al. (1999a,b, 2001) and we tabulate those values relevant to our analysis
in table 1 of M01a.

The values of $\omega_0$ must be known to a high precision in order to
place stringent constraints on $\da$.  A precision of $\da \sim 10^{-5}$
corresponds to an uncertainty on $\omega_0$ $\sim0.02{\rm \,cm}^{-1}$
($\sim 0.3 {\rm \,kms}^{-1}$)\footnote{ We can see this as follows.  We can
ignore the term in $q_2$ since these coefficients are typically an order of
magnitude smaller than the $q_1$ coefficients.  Consider two transitions,
one with a low atomic mass, the other a high atomic mass.  Typically, the
difference between the $q_1$ coefficients for a light species (e.g. Mg{\sc
\,ii}) and a heavy species (e.g. Fe{\sc \,ii}) is $\sim1000{\rm
\,cm}^{-1}$.  Equation \ref{eq:omega}, applied to these two transitions,
then shows that a precision of $\omega_0$ of $\sim0.02{\rm \,cm}^{-1}$ is
required in order to detect a variation of $\da \sim 10^{-5}$.}.  This
corresponds fairly well to the data quality since we can centroid an
absorption feature to approximately one tenth of the HIRES resolution,
$\sim 7{\rm \,kms}^{-1}$.  Such precision has not been previously achieved
for most resonance transitions of ionized species and so measurements of
$\omega_0$ for all relevant species were carried out by several groups
(Nave et al. 1991; Pickering, Thorne \& Webb 1998; Pickering et al. 2000;
Griesmann \& Kling 2000) using laboratory Fourier transform
spectrometers. We have summarised and tabulated these measurements in M01a.

Knowing all values of $\omega_0$, $q_1$ and $q_2$, we then deconvolve each
absorption system into individual Voigt components and perform a non-linear
least squares $\chi^2$ minimization to find the best-fitting value of $\da$
at each absorption redshift. For the \loz~sample we confirm the W99
result, our new results giving a weighted mean of $\da = (-0.70 \pm
0.23)\times 10^{-5}$. For the new, \hiz~sample, we find a similar
result: $\da = (-0.76 \pm 0.28)\times 10^{-5}$. Both results independently
suggest a smaller value of $\alpha$ in the past. Taking the sample as whole
we find $\da = (-0.72 \pm 0.18)\times 10^{-5}$, a $4\sigma$ result. These
new results are also summarised in Webb et al. (2001, hereafter W01).

Given the potential importance of such a result for fundamental physics, a
thorough investigation of possible systematic problems in the data or in
the analysis is essential, and this is the purpose of the present paper. In
Section 2 we list all effects that one might consider as being able to
mimic a systematically non-zero $\da$. We are able to exclude many of them
with general arguments. Sections 3--7 are then given over to a more
detailed analysis of five specific effects: wavelength mis-calibration,
ionic line blending, atmospheric dispersion effects, differential isotopic
saturation and possible isotopic variation. Section 8 describes another
test which is sensitive to a simple monotonic wavelength distortion,
specifically for the \hiz~sample. We make our conclusions in
Section 9.

\section{Summary of potential systematic errors}
In M01a and W01 we found that values of $\da$ were systematically negative
and that this trend was evident in two different samples of data occupying
different redshift regimes. We note that the \loz~sample is
more susceptible to systematic errors than the \hiz~sample. The
Mg{\sc \,ii} lines act as anchors against which the larger Fe{\sc \,ii}
shifts can be measured.  All of the Fe{\sc \,ii} lines have similar
magnitude shifts in the same direction (i.e. $q_1$ is large and positive in
all cases). Furthermore, all the Fe{\sc \,ii} lines lie to the blue of the
Mg{\sc \,ii} doublet.  If some systematic error in the wavelength scale is
present, which effectively {\it compresses} the spectrum, this would mimic
a negative shift in the mean value of $\da$ (if it were not degenerate with
redshift). The \hiz~sample is far more complex in this regard
since it contains ions not only with positive coefficients, but also with
negative $q_1$ coefficients (i.e. all Cr{\sc \,ii} lines and Ni{\sc \,ii}
$\lambda$1741 and $\lambda$1751). Also, the shifts for the various ions are
of various magnitudes making it more complicated to predict the effect of a
systematic error on our measured values of $\da$.

\subsection{Laboratory wavelength errors}\label{sec:laberrors}
Errors in the values of $\omega_0$ will lead directly to errors in any
single determination of $\da$. If, for example, the laboratory wavelengths
of the Mg{\sc \,ii} lines were all shifted to the blue or red then this
would systematically bias the values of $\da$ for the \loz~sample. In M01a
we summarise the measurements of all values of $\omega_0$.  Most
measurements were repeated independently and good agreement was obtained in
those cases.  A typical measurement accuracy from these laboratory
measurements $\omega_0$ of is $\sim 2 \times 10^{-3}{\rm \,cm}^{-1}$, which
could only introduce a maximum error in $\da$ around an order of magnitude
below the observed deviation from zero.  In practice, since we use several
transitions (particularly for the \hiz~sample), any associated errors are
likely to be smaller.  We therefore consider it unlikely that the
systematic shift in $\da$ is due to errors in the values of $\omega_0$.

As also noted in M01a, some laboratory spectra were calibrated with the
Ar{\sc \,ii} lines of Norl\'{e}n (1973) while others were calibrated with
those of Whaling et al. (1995). The Whaling wavenumbers are systematically
larger than those of Norl\'{e}n such that $\Delta\omega = 7\times
10^{-8}\omega$. To deal with this discrepancy we have re-scaled those
wavenumbers which had been calibrated using the Whaling scale (Al{\sc
\,ii}, Al{\sc \,iii}, Si{\sc \,ii}) to the Norl\'{e}n scale.  We stress
that the difference between the Norl\'{e}n and Whaling wavenumbers is a
{\it scaling only}, and not a distortion.  Therefore, is it not possible to
introduce an apparent deviation in $\da$ from zero, since small zero-point
offsets in the wavelength scale are degenerate with redshift.

\subsection{Heliocentric velocity variation}\label{sec:helio}
During a one hour exposure, representative of the quasar integrations
for the sample we have used, the heliocentric velocity may
change by as much as $\sim 0.1{\rm \,kms}^{-1}$. This will act to
smear any spectral feature: the instantaneous spectrum is convolved
with a top hat function in velocity space. However, since we fit a
redshift parameter to the spectrum when determining $\da$ then a
velocity space smearing is completely absorbed into the redshift
parameter. Heliocentric smearing will have no effect on measurements
of $\da$.

\subsection{Differential isotopic saturation}\label{sec:sat}
The ions used in our analysis, with the exception of those of Al, have
naturally occurring isotopes. Thus, each absorption line will be a
composite of absorption lines from all the isotopes of a given
species. In all cases other than Mg and Si, the laboratory wavenumbers
we have used correspond to {\it composite} values.  The composite
wavenumbers will only strictly be applicable in the optically thin
regime (linear part of the curve of growth).  As the column
density increases, the strongest isotopic component begins to saturate
and the line centroid will shift according to the natural abundances
(cf. Section \ref{sec:iso}) of the other isotopes.

The mass isotopic shift, being the dominant splitting mechanism for light
atoms, is proportional to $\omega_0/m^2$ for $m$ the atomic mass.  The
wavelength separation between the isotopes is therefore largest for the
lighter (low $q_1$) ions.  Isotopic separations have been measured for the
Mg{\sc \,i} $\lambda$2853 transition by Hallstadius (1979), for Mg{\sc
\,ii} $\lambda$2796 by Drullinger et al. (1980) and the isotopic structure
of all Mg lines is presented in Pickering et al. (1998).  However,
similar measurements do not exist for the other transitions of
interest. For Si we used estimates of the isotopic wavenumbers based on a
scaling from the basic Mg structure by the mass shift.  For all other
species, measured (composite) wavenumbers were used.  See M01a for full
details.

We discuss and quantify this effect in Section \ref{sec:saturation}.

\subsection{Isotopic abundance variation}\label{sec:iso}
In the above we have assumed that the isotopic abundances in the absorption
systems are equal to terrestrial values (Table \ref{tab:iso}). However, if
the isotopic abundances at high redshift are significantly different then
this may lead to apparent shifts in the absorption line centroids and this
would cause a systematic error in $\da$. Again, the most susceptible
transitions are those of Mg and Si since (i) these ions will have the
largest isotopic separations (i.e. $m$ is low) and (ii) transitions in
these atoms are prominent in our data.  If, for the \loz~sample,
the $^{25}$Mg and $^{26}$Mg abundances were zero, we would incorrectly
infer a more positive $\da$ by assuming terrestrial isotopic abundances.
In the simple case of finding $\da$ from Mg{\sc \,ii}\,$\lambda$2796 and
Fe{\sc \,ii}\,$\lambda$2344, we would expect to find $\da \approx 0.6
\times 10^{-5}$ in the absence of any real variation in $\alpha$.

We consider this possibility in detail in Section \ref{sec:noiso}

\subsection{Hyperfine structure effects}\label{sec:hyper}
Hyperfine splitting of energy levels occurs in species with odd proton or
neutron numbers. Different hyperfine components will have different
transition probabilities but the composite wavelength of a line will be
unchanged by the splitting (i.e. the centre of gravity of the hyperfine
components is constant with increased splitting). However, a similar
differential saturation effect will occur for the hyperfine components as
discussed in Section \ref{sec:sat} for the isotopic components.

The worst case here is that of $^{27}$Al -- the only stable isotope of Al
-- since it has an odd proton number and a very large magnetic moment. In
the laboratory experiments of Griesmann \& Kling (2000), the hyperfine
structure of the Al{\sc \,iii} doublet was clearly resolved and can be seen
to be quite prominent. However, as noted in M01a, we only include Al{\sc
\,iii} lines in the analysis of 6 absorption systems of our \hiz~sample and
in these cases we see very little or no saturation. The effect of removing
the individual Al{\sc \,iii} transitions from those systems on $\da$ can be
seen in Fig. 4.  In the case of Al{\sc \,ii} $\lambda$1670, both the ground
and excited state have zero spin and the $p$-wave hyperfine splitting in
the excited state is very small (cf. $^{25}$Mg in Pickering et
al. 1998). Thus, we do not expect any Al hyperfine saturation effects on
$\da$.

All other dominant isotopes in our analysis have even proton numbers and
where odd isotopes occur these have very low abundances (see Table
\ref{tab:iso}). The $^{25}$Mg hyperfine structure has already been taken
into account in our analysis but the hyperfine separations are so small
that neglecting it would make no difference to $\da$ even upon
saturation. The odd isotopes of Si, Cr, Fe, Zn and Ni have very low
abundances. Also, Si, Cr, Fe and Ni will have very small hyperfine
splitting due either to low magnetic moments or to the nature of the ground
and excited state wavefunctions.

We therefore expect that differential hyperfine structure saturation has
not effected $\da$ in our analysis. An unlikely interpretation, given the
above, is cosmological evolution of the hyperfine structure constants but
very large relative variations would be required.

Another possibility is that the populations of the two hyperfine levels of
Al{\sc \,iii} are not equal in the absorption clouds. Since the gas clouds
have very low density, the equilibrium between the two hyperfine levels
will be maintained predominantly by interaction with cosmic microwave
background (CMB) photons. For the Al{\sc \,iii}\,$\lambda$1862 transition,
this implies a lower limit on the relative populations of the two levels of
$\exp(-\Delta\omega/k_{\rm B}T_{\rm CMB}) \approx 0.9$, leading to a shift
in the line centroid of $\sim 0.01{\rm cm}^{-1}$ ($\sim 5\times 10^{-4}{\rm
\,\AA}$). This is a potential problem for the 6 systems in which we fit
Al{\sc \,iii} lines. We do not consider this effect in detail but we note
that Fig. 4 in Section \ref{sec:linrem} shows that potential lack of
thermal equilibrium has no significant effect on $\da$.

\subsection{Magnetic fields}\label{sec:magfields}
Shifting of ionic energy levels due to large magnetic fields may also be a
possibility. If very large scale magnetic fields exist in the intergalactic
medium and quasar light is polarized, then this could result in correlated
values of $\da$ in neighbouring regions of the universe. However, the QSOs
studied in M01a and W01 cover a large region of the sky.  Furthermore, even
in Abell clusters, intra-cluster magnetic field strengths are typically
$\sim \mu$G (e.g. Feretti et al. 1999), roughly 9 orders of magnitude below
the strength required to cause substantial effects.  We consider this
possibility to be very unlikely.

\subsection{Kinematic effects}\label{sec:dynamics}
An assumption in the analysis of M01a and W01 is that the absorption
velocity structure is the same in different species.  Of course, if there
were some non-zero velocity between, say, the Mg{\sc \,ii} and Fe{\sc \,ii}
components of a given cloud then we would incorrectly determine $\da$.
However, it is difficult to imagine a mechanism by which the Mg{\sc \,ii}
is systematically blue or redshifted with respect to the Fe{\sc \,ii} when
averaged over a large sample of absorption systems.  Suppose for example we
have an asymmetric, expanding (or contracting), rotating cloud in which all
the Fe{\sc \,ii} and Mg{\sc \,ii} lie at different `radii'. Any large
sample of sight lines through such a cloud would yield an average value of
$\da = 0$.  Furthermore, if this was a major effect then the observed
scatter in the $\da$ points would be greater than expected on the basis of
our $\da$ error estimates, and in M01a we showed this not to be the case.
No matter how contrived the velocity structure and ionic segregation, when
averaging over a large number of values of $\da$, it is not feasible to
generate a systematic offset from zero.

\subsection{Wavelength mis-calibration}\label{sec:wavemiscal}
There may be inaccuracies in the wavelength calibration software used
to reduce the spectra. Also, human error cannot be ruled out.  W99
stated that a drift in the wavelength calibration drift by an amount
corresponding to roughly $\sim 2.5$ times the mean wavelength-pixel
residuals over the range of the CCD could result in a significant
error in the \loz~results. A considerably more complicated
form of the mis-calibration would be required to produce a systematic
error in the \hiz~results.

The wavelength calibration of the QSO CCD images is done by comparison
with Thorium--Argon (ThAr) lamp exposures taken before and after the
QSO frames. We consider errors in the laboratory wavelengths of the
selected ThAr lines to be negligible as their relative values are
known to much greater precision than our values of $\omega_0$ (Palmer
\& Engleman Jr. 1983).  However, ThAr line mis-identifications
may lead to serious mis-calibration of the wavelength scale over large
wavelength regions. Such mis-identifications can also be applied to the
rest of the spectra taken over the same series of observations
(i.e. applied to the spectra of many QSOs). Thus, {\it a priori}, we
cannot rule out the possibility that the wavelength scale has been
set improperly in this process and that a systematic shift in $\alpha$
has not been mimicked.  Such a potential effect needs careful
investigation.

In Section \ref{sec:wavcal} we discuss and quantify
this effect for our data set.

\subsection{Air--vacuum wavelength conversion}\label{sec:airvac}
Most ThAr line lists are presented as air wavelengths.  The usual
data reduction procedure is to carry out the wavelength calibration
fits to the data using air wavelengths and then convert the calibrated
spectrum to vacuum wavelengths.
For example, {\sc \,iraf}\footnote{{\sc iraf}
is distributed by the National Optical Astronomy Observatories, which are
operated by the Association of Universities for Research in Astronomy,
Inc., under cooperative agreement with the National Science Foundation.}
uses air wavelengths for a selection of lines from the Thorium spectral
atlas of Palmer \& Engleman Jr. (1983) and from the Argon lines of
Norl\'{e}n (1973). The same list is used in {\sc makee} -- the HIRES data
reduction package written by T. Barlow. {\sc makee} was used to calibrate
12 of our \hiz~sample and the remainder were calibrated within
{\sc iraf}. The latter were actually calibrated with a set of vacuum
wavelengths, $\lambda_{\rm vac}$, that {\sc iraf} states is derived from
the air wavelengths, $\lambda_{\rm air}$, by using the Edl\'{e}n (1966)
formula for the refractive index, $n$,
\begin{equation}\label{eq:edlen}
10^8\left[n(\lambda_{\rm vac})-1\right]=8342.13 + \frac{2406030}{(130 -
\sigma^2)} + \frac{15997}{(38.9 - \sigma^2)}
\end{equation}
at $15^{\circ}{\rm C}$ and atmospheric pressure. Instead of using $\sigma
\equiv 10^4/\lambda_{\rm vac}$, {\sc iraf} makes the approximation, $\sigma
\approx 10^4/\lambda_{\rm air}$. {\sc makee} converts the final air
wavelength calibrated QSO frames to vacuum using the Cauchy dispersion
formula (Weast 1979),
\begin{equation}\label{eq:cauchy}
10^7\left[n(\lambda_{\rm air})-1\right]=2726.43 +
\frac{12.288}{10^{-8}\lambda_{\rm air}^2} +
\frac{0.3555}{10^{-16}\lambda_{\rm air}^4}
\end{equation}
at $15^{\circ}{\rm C}$ and atmospheric pressure. $\lambda_{\rm vac}$ and
$\lambda_{\rm air}$ are both measured in \AA ngstroms in the above
equations. This difference begs the question, what is the absolute and
relative accuracy of these conversion formula?

First let us refine this question. The measurements reported by Palmer \&
Engleman Jr. and Norl\'{e}n were all carried out in vacuum. The air
wavelengths they quote are calculated using the Edl\'{e}n formula. Thus,
{\sc iraf} and {\sc makee} will only produce `correct' vacuum wavelength
scales if they invert the Edl\'{e}n formula to calculate the refractive
index. As stated above, this is not the case. Therefore, to re-state the
above question, what error does the {\sc iraf} and {\sc makee} air--vacuum
wavelength conversion introduce into the resulting QSO wavelength scale? To
illustrate the answer to this we plot the various conversion formulae
commonly used in the literature, including those discussed above, in
Fig. 1. We have also included the original Edl\'{e}n (1953)
formula\footnote{The Edl\'{e}n (1953) formula is of the same form as the
1966 formula (equation \ref{eq:edlen}) but has slightly different parameters:
6432.8, 2949810, 146, 25540 and 41 in order from left to right, numerator
to denominator.} for comparative purposes. Peck \& Reeder (1972) published
a fit to revised refractive index data, proposing a 4-parameter formula:
\begin{equation}\label{eq:peck}
10^8\left[n(\lambda_{\rm vac})-1\right]=\frac{5791817}{(238.0185 -
\sigma^2)} + \frac{167909}{(57.362 - \sigma^2)}
\end{equation}
at $15^{\circ}{\rm C}$ and atmospheric pressure with $\sigma\equiv
10^4/\lambda_{\rm vac}$ (again, $\lambda_{\rm vac}$ is in \AA
ngstroms).

\begin{figure}
\label{fig:compedlen}
\centerline{\psfig{file=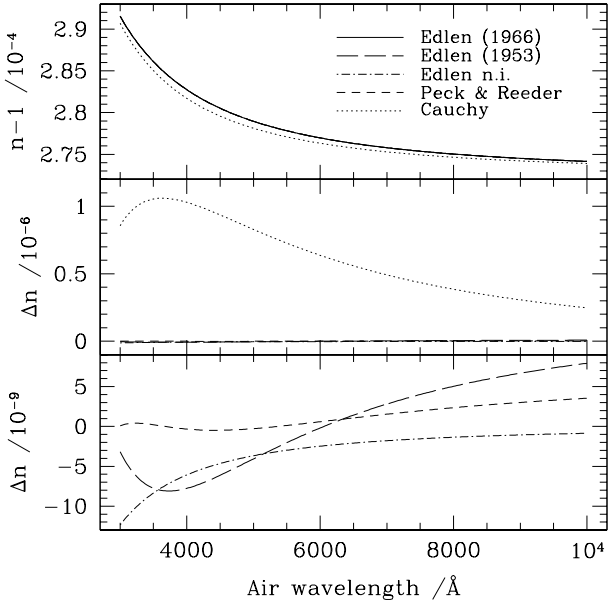,width=8.5cm}}
\caption{Comparison of the various air dispersion formulas used in the
literature (all for $15^{\circ}{\rm C}$ and atmospheric pressure). We use
the Edl\'{e}n (1966) formula as our reference since the atlases of Palmer
\& Engleman Jr. (1983) and Norl\'{e}n (1973) use this to convert their
vacuum wavelengths to air wavelengths. The top panel shows the dispersion
for the Edl\'{e}n (1966) and Edl\'{e}n (1953) formulas together with that
of Peck \& Reeder (1972) and the Cauchy formula (Weast 1979). We also plot
the approximation made within {\sc iraf} where the Edl\'{e}n formula has
not been inverted to convert air wavelengths to vacuum (denoted ``Edlen
n.i.''). The middle and lower panels show the difference between the
Edl\'{e}n (1966) refractive index, $n_{\rm E}$, and that derived from the
other formulae, $n_{\rm X}$: $\Delta n \equiv n_{\rm E} - n_{\rm X}$.}
\end{figure}

From Fig. 1 we see that the Cauchy formula seriously deviates from the
other formulas. Furthermore, the difference between the Edl\'{e}n and
Cauchy dispersions is strongly wavelength dependent. Thus, spectra
calibrated with {\sc makee} may have systematically non-zero values of
$\da$ due to this distortion. As we stated in M01a, we have corrected the
12 effected spectra in our (\hiz) sample by converting the wavelength scale
back to air wavelengths using the (inverted) Cauchy formula and then back
to vacuum using the (inverted) Edl\'{e}n formula. From Fig. 1 we can also
see that there is only a small distortion to the wavelength scale
introduced due to the approximation made within {\sc iraf}. This difference
is below our level of precision.

\subsection{Temperature changes during observations}\label{sec:temp}
The refractive index of air within the spectrograph depends on
temperature (and also on pressure but this is a smaller effect). If
a QSO spectrum is calibrated with only a single ThAr exposure taken at
a different spectrograph temperature, we expect a systematic
mis-calibration of the QSO frame's wavelength scale. Consider a QSO
frame taken at $15^{\circ}{\rm C}$ which is calibrated with a ThAr
exposure taken at $0^{\circ}{\rm C}$ and similar pressure. Any
wavelength separation in the QSO exposure will be overestimated. For
example, the difference between the wavelength separations of vacuum
$\lambda4000$ and $\lambda7000{\rm ~\AA}$ at $15^{\circ}{\rm C}$ and
$0^{\circ}{\rm C}$ is $43\times 10^{-3}{\rm ~\AA}$ (using the tables
in Weast 1979). A comparison with values of $\Delta \lambda$ in table
1 of M01a shows that this would result in a significantly positive
$\da$ for the \loz~sample, the situation not being so clear
for the \hiz~sample.

Note that the above effect can only mimic a systematically non-zero
$\da$ if the spectrograph temperature is systematically higher or
lower for the ThAr frames compared with the QSO frames. This is
possible if the ThAr exposures were always taken before or after the
QSO frames and the temperature evolves monotonically throughout the
night. However, the effect is greatly reduced if ThAr exposures are
taken near the time of the QSO observations. This was the case in our
observations. We have used image header information to calculate the
QSO--ThAr temperature difference, $\Delta T \equiv \left<T_{\rm
QSO}\right> - \left<T_{\rm ThAr}\right>$, inside HIRES for both the
low and \hiz~samples. Here the average of $T_{\rm QSO}$ and
$T_{\rm ThAr}$ is taken over all exposures for each object. We find
mean values of $\Delta T = 0.04 \pm 0.02{\rm ~K}$ and $\Delta T = 0.2
\pm 0.1{\rm ~K}$ for the low and \hiz~samples
respectively. Taking into account that, on average, rest-frame
separations between transitions are $\la 300{\rm ~\AA}$ then we can
see that temperature variations could not have mimicked any
significant shift in $\da$ in either sample.

\subsection{Line blending}\label{sec:blending}
The errors in $\da$ presented in M01a and W01 take into account errors from
signal-to-noise and spectral resolution considerations and the
velocity structure of the profile fits. The errors are also reduced
when more lines are fitted simultaneously. However, we have assumed
that we have deconvolved each absorption system into the correct
number of velocity components. There may have been weak, interloping,
unresolved lines which, if the interloping species were in the same
absorption cloud, could have produced a shift in the fitted line
wavelengths of all velocity components of one or more transitions.

We distinguish between {\it random} blends and {\it systematic}
blends. {\it Random} blends may occur if many absorption clouds at
different redshifts intersect the line of sight to a single QSO.  A
{\it systematic} blend will occur when two species are in the same
cloud and have absorption lines with similar rest-wavelengths.  Such
an effect could mimic a systematic shift in $\alpha$.

The importance of such an effect diminishes as the number of transitions
used in each fit is increased.  Therefore, the effect, if present, would be
expected to be smaller in our \hiz~sample.

In Section \ref{sec:sysblend} we describe the results of a detailed
search for atomic (and not molecular) interlopers for all transitions
of interest, and place constraints on their strengths and positions.

\subsection{Atmospheric dispersion effects}\label{sec:atmodisp}
As noted in M01a, all of the \loz~sample and many of the high redshift
sample of QSOs were observed before 1996 August at which time an image
rotator was fitted to HIRES.  Objects observed prior to August 1996 were
therefore not observed with the slit length (spatial direction)
perpendicular to the horizon.  Atmospheric dispersion leads to a stretching
of the target spectrum relative to the calibration spectrum.

An additional consequence is the potential wavelength-dependent
truncation of the QSO seeing profile on the slit jaw edges.  This can
introduce a wavelength dependent asymmetry in the point-spread
function (PSF).

Together, these effects can conspire to mimic a non-zero $\da$.  We provide
detailed calculations for these effects in Section 5 and show that they
will tend to produce an apparent {\it positive} $\da$, and so cannot mimic
our results.

\subsection{Instrumental profile variations}\label{sec:IP}
If the instrumental profile (IP) of HIRES shows significant {\it
intrinsic} asymmetries then we expect absorption line centroids to be
incorrectly estimated.  If the asymmetry varies with wavelength, this
could mimic a non-zero $\da$.  Valenti, Butler \& Marcy (1995) have
determined the HIRES IP for several positions along a single order and
they find that the IP is indeed asymmetric and that the asymmetry
varies slightly along the echelle order. Valenti et al. (1995), do not
quantify the asymmetry variation across orders although it is likely
to be comparable to the variation along the orders.

As a by-product of the detailed investigation into possible wavelength
calibration errors, we find that this effect is negligible (see Section
\ref{sec:wavcal} for details).

\bigskip

We have therefore eliminated, with some reliability, potential
systematic errors due to: laboratory wavelength errors, heliocentric
velocity variation, hyperfine structure effects, magnetic fields,
kinematic effects, air--vacuum wavelength conversion errors and
temperature changes during observations.

In the following sections we investigate in considerable detail the
remaining effects: wavelength mis-calibration (and instrumental profile
variations), line blending, atmospheric dispersion effects,
differential isotopic saturation and isotopic abundance variation.  We
also consider a simple test for any simple but unidentified systematic
errors for the \hiz~sample in Section \ref{sec:apns}.

\section{Detailed analysis: wavelength calibration errors}\label{sec:wavcal}
In this section, we investigate the possibility that wavelength calibration
errors could have lead to an apparent non-zero $\da$.  To quantify this
directly, we analyse sets of ThAr emission lines in the calibration lamp
spectra in the same way as each set of QSO absorption lines has been
analysed.  If no calibration error is present then we expect that
$(\da)_{\rm ThAr} = 0$ for all clouds (i.e. for all $z$).

\subsection{Method}

For each QSO spectrum, we selected $\sim$15\,\AA\, sections of ThAr
spectra at wavelengths corresponding to the observed absorption lines
in the QSO spectra.  Each ThAr $\sim$15\,\AA~section contains several
(typically up to $\sim 10$) lines.  Error arrays were generated
assuming Poisson counting statistics.  We fitted continua to these
ThAr sections, dividing the raw spectra by the continua to obtain
normalized spectra.

The main steps in the procedure are then as follows.  We select
several independent sets \footnote{Each set of ThAr lines never
included any of the ThAr lines from any other set.} of ThAr lines.
Each set of ThAr lines therefore corresponds to, and has been selected
in an analogous way to, one QSO absorption system.  

In order to obtain the best estimate of any wavelength calibration
errors, the average $(\da)_{\rm ThAr}$ is determined for each ThAr set
(typically 3--7 sets for each QSO absorption system).  The process of
defining sets in this way also provides a direct additional estimate
of the errors on $(\da)_{\rm ThAr}$.

To fit the ThAr spectra, a modified version of {\sc vpfit} was used (to fit
Gaussian profiles, see M01a), adopting ThAr laboratory wavelengths from
Palmer \& Engleman Jr. (1983) for values of $\omega_0$ in equation
\ref{eq:omega}. The $q_1$ and $q_2$ coefficients for the QSO absorption
lines are applied to the corresponding ThAr lines. That is, we treat the
set of ThAr lines as if they were the QSO lines.  The free parameters
involved in each fit were line width (assumed the same for all ThAr lines
in a given set, although the results were insensitive to this assumption),
`redshift' (numerically close to zero) and peak intensity.

\subsection{Results and Discussion}
We applied the above method to each available ThAr spectrum
corresponding to each QSO absorption cloud considered and the results
are illustrated in Fig. 2. Our first observation is that the scatter
in these results is an order of magnitude less than that of the QSO
absorption results themselves.  This clearly demonstrates that
wavelength mis-calibration of the CCD has not mimicked a shift in
$\alpha$ to any significant degree. The results above also show that
the data quality permit a precision of $\da \sim 10^{-7}$.

\begin{figure}
\label{fig:wavecal}
\centerline{\psfig{file=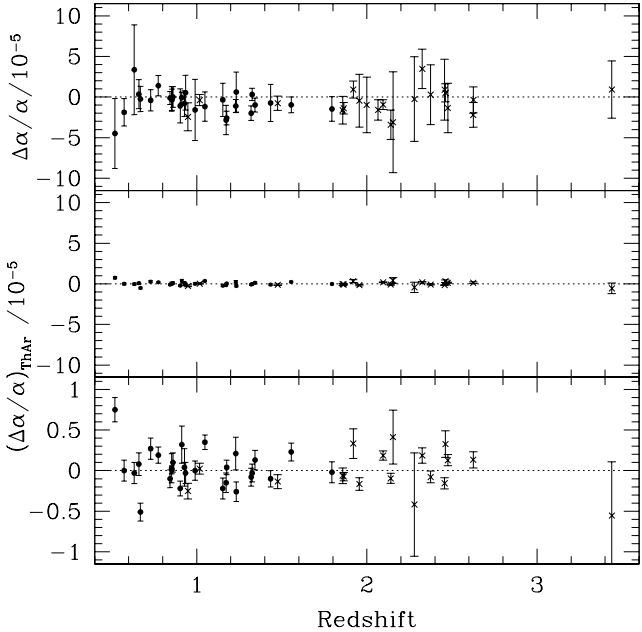,width=3.5in}}
\caption{The top panel shows the results of M01a and W01. The middle panel
compares the ThAr calibration error on the same scale. The lower panel
gives an expanded view of the ThAr results. The weighted mean of the
ThAr sample is $(0.0 \pm 1.4) \times 10^{-7}$. ThAr data were not
available for five of the absorption systems. Clearly, this does not
effect our conclusions.}
\end{figure}

Secondly, many points in Fig. 2 deviate significantly from zero.
There are three possible reasons for this: weak blended emission
lines, ThAr line mis-identifications, and errors in the ThAr laboratory
wavelengths.  Inspecting the ThAr spectra does indeed reveal typically
$\sim$20 easily identifiable, but very weak, lines per $15$\,\AA\, and
we would expect this to add to the scatter in the final results for
$(\da)_{\rm ThAr}$.  Furthermore, this effect would not produce
correlated deviations in $(\da)_{\rm ThAr}$, as observed.  Consistent
with this interpretation, we found that the value of $\chi^2$ per
degree of freedom for a fit to a given set of ThAr lines was $\gg 1$.
Note that if ThAr line mis-identifications had occurred, one may expect
systematically correlated deviations.  Finally, we do not expect any
significant errors in the ThAr laboratory wavelengths (as discussed in
Section \ref{sec:wavemiscal}).

On inspection of some of the ThAr spectra, we noted that some ThAr
lines towards the blue end of the spectrum were slightly asymmetric.
This may indicate a degradation in the polynomial fits near the edges
of the fitting regions.  There may also be intrinsic asymmetries in
the IP (Section \ref{sec:IP}). If such asymmetry variations exist then
they should also apply to the QSO spectra.  However, the results in
Fig. 2 show that such variations do not produce a systematic effect in
$\da$ at any detectable level.

We therefore conclude this section by stating that the results in
Fig. 2 show unambiguously that wavelength calibration errors and
variations in IP asymmetries are not responsible for the observed
shifts in $\alpha$ in the QSO absorption line results.

\section{Detailed analysis: systematic line blending with unknown 
species}\label{sec:sysblend}

In this section we suppose there is an unidentified transition, arising
from some species in the same gas cloud as is being studied.  We further
assume that this interloping transition is not strong enough and/or
sufficiently displaced from one of the lines in the analysis, to be
detected directly.  We describe two approaches to explore this: in Section
4.1 we attempt to identify candidate blending species and in Section 4.2 we
describe the results of a test where we remove one transition or one
species at a time to investigate the impact on $\da$.

\subsection{Search for potential interlopers}

There are two main parameters which will determine whether or not any
particular interloper can give rise to a significant effect in terms of
$\da$: column density and the position in wavelength with respect to
the `host' line of interest.  In order to search for a potential
interloper, we explore the range in parameter space through numerical
simulation, generating trial interlopers, blended with a single
stronger `host' line, and examining the resulting shift in the
combined centroid in terms of $\da$.  Having established the allowed
characteristics of a potential interloper, we use photoionization
equilibrium models to search for candidate atomic species.  We ignore
molecular species.

\subsubsection{Characteristics of a candidate interloper}
We restrict the following discussion to blending with Mg{\sc \,ii}
$\lambda$2796, since $\da$ is particularly sensitive to blending with this
anchor line in the \loz~sample. We later generalize our discussion to
include blending with all other relevant transitions.

Consider an interloper -- a spectral line of a blending species $X$ --
which is separated from the fitted line centroid wavelength,
$\lambda_0$, by $-\Delta\lambda$ ($\Delta\lambda > 0$) in the frame of
the cloud. Let the separation between the actual Mg{\sc \,ii}
$\lambda$2796 wavelength and the fitted line position be $d\lambda$
(i.e. $d\lambda > 0$) in the frame of the cloud. That is, the line is
fit as a single component, despite the presence of the interloper. The
interloper has a column density $N(X)$. However, as the species $X$ is
not known {\it a priori}, we may only determine the quantity $N(X)f_X$
where $f_X$ is the oscillator strength of the interloping transition.

We generated a synthetic optically thin Mg {\sc \,ii} $\lambda$2796 line
and matched this against a composite Mg {\sc \,ii} $\lambda$2796 profile
which included an additional weak blended component, representing our
candidate interloper.  The model was fitted to the `data' using {\sc
vpfit}.  The synthetic spectrum was generated using a column density of $4
\times 10^{12}{\rm ~cm}^{-2}$ and a $b$-parameter of $5.0{\rm ~kms}^{-1}$
as these were representative values for the Mg{\sc \,ii} lines in M01a and
W01 (Churchill 1997).

We restricted the range of $b$-parameters for the interloper using the
following physical considerations. If we consider all lines to be thermally
broadened then the $b$-parameter varies inversely as the square root of the
atomic mass. Any species lighter than Mg will have a larger
$b$-parameter. Since we are considering $X$ to have transitions in the
ultraviolet, we may assume that it probably has a mass greater than or
comparable to that of Mg. We therefore place an upper limit of $b(X) <
6.0{\rm ~kms}^{-1}$. Also, we did not consider any species with atomic
number greater than $\sim$100 and so we adopted a lower limit of $b(X) >
0.5{\rm ~kms}^{-1}$.

By comparing sets of model and synthetic blended lines in this way, we
can constrain the range of possible values of $d\lambda$, given some
apparent $\da$.

We present the results of the simulation in Fig. 3 where we define
$\Gamma_{X,{\rm Mg}{\sc \,ii}} \equiv \log_{10}[N(X)/N({\rm Mg}{\sc
\,ii})]$ and $F_X \equiv \log_{10}f_X$. The ripples on the three surface
reflect errors ($\la 0.1{\rm ~dex}$ in $\Gamma_{X,{\rm Mg}{\sc \,ii}}$)
introduced due to the restrictive upper limit placed on $b(X)$.
\begin{figure}
\label{fig:Nplot}
\centerline{\psfig{file=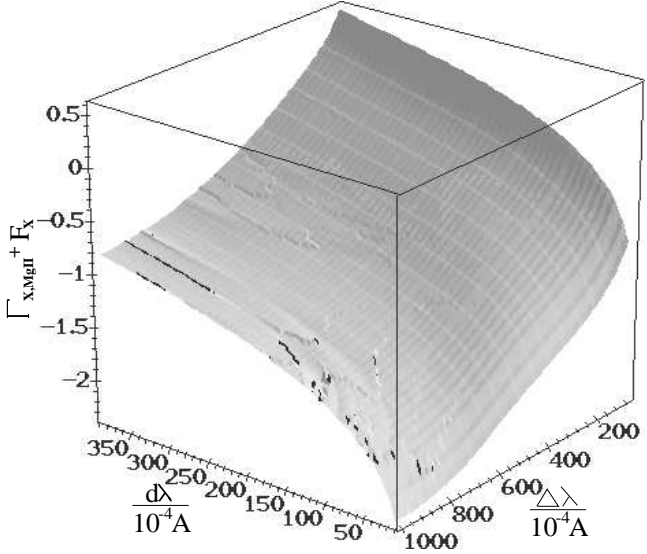,angle=270,width=8.5cm}}
\caption{Three space of results from the blending simulation. $d\lambda$
directly corresponds to a value of $\da$ if blending is responsible for all
observed shifts in figure 3 of M01a. The figure is shaded according to
values of $\Gamma_{X,{\rm Mg}{\sc \,ii}}$. Ripples on the surface reflect
errors due to the restrictive upper bound on $b(X)$. The `lighting' of the
figure has been adjusted to maximize their visibility and the main errors
have been highlighted in black.}
\end{figure}

Using Fig. 3, we can now specify a certain value of $\da$ (i.e. we specify
$d\lambda$) and an interloping transition of a species $X$ (i.e. we specify
$\Delta\lambda$) and find the lower bound on the column density which that
species must have in order to have caused the said shift in $\da$. For
example, if we are concerned with $\da \sim 10^{-5}$ then $d\lambda \sim 50
\times 10^{-4}$\AA\, and the section of Fig. 3 with $d\lambda \la 50
\times 10^{-4}$\AA\, contains information which restricts the candidate
interlopers.

\subsubsection{Photoionization Modeling}
Given a species, $X$, what is an upper limit on the column density in
any given cloud? To answer this, we modelled the absorption clouds
with a grid of {\sc cloudy} models (Ferland 1993). {\sc cloudy}
estimates the photoionization conditions in absorption clouds by
assuming that the cloud can be approximated as a series of zones, in a
plane parallel geometry, illuminated by UV radiation from one
direction.  The fact that we are trying to model a (presumably)
spherical cloud with UV flux incident from all sides is expected to
introduce abundance errors of order a factor of 2 (Ferland 1993).

Photoionization equilibrium is characterized by (a) the shape and intensity
of the UV spectrum, (b) the number density of hydrogen atoms in the cloud,
$n_{\rm H}$, (c) the size of the cloud, which we assume is the same for all
species (the size can be found by specifying $n_{\rm H}$ and $N({\rm H}{\sc
\,i})$, the neutral hydrogen column density) and (d) the metallicity, $Z$,
of the cloud. It is also useful to define the ionization parameter as $U
\equiv n_{\gamma}/n_{\rm H}$, where $n_{\gamma}$ is the number density of
ionizing photons incident on the cloud.  Here we follow the work in
Bergeron \& Stasi\'{n}ska (1986), Bechtold et al. (1987), Steidel \&
Sargent (1989) and Bergeron et al. (1994) and choose a power law spectrum,
$f_{\nu} \propto \nu^{-3/2}$, for $f_{\nu}$ the incident flux at frequency
$\nu$. The low and high energy extremes are given by $f_{\nu} \propto
\nu^{5/2}$ (for $\lambda_{\nu} > 10~\mu{\rm m}$) and $f_{\nu} \propto
\nu^{-2}$ (for $E_{\nu} > 50{\rm ~keV}$).

All the above parameters ($U$, $n_{\rm H}$, $N({\rm H}{\sc \,i})$ and $Z$),
vary from cloud to cloud and are poorly known quantities. We therefore use
a grid of cloudy models to find an upper limit on the column density of a
given ionic species using ranges in these parameters as follows. For $U$ we
adopt the range $-4 < \log_{10}U < -3$ in line with the measured range of
parameters for those clouds studied in M01a and W01 (Churchill
1997). Similarly, for $N({\rm H}{\sc \,i})$ we use $17.0 < \log_{10}N({\rm
H}{\sc \,i}) < 19.0$. For $Z$ we adopt a conservative range, $-2< Z < 0$
(following Morris et al. 1986; Bergeron \& Stasi\'{n}ska 1986; Bechtold et
al. 1987; Steidel \& Sargent 1989; Bergeron et al. 1994; Churchill
1997). The range of parameters we explore here are characteristic of the
low redshift sample. For a related discussion of DLAs, see Prochaska (1999)
and Prochaska \& Wolfe (2000).

$n_{\rm H}$ is the most difficult parameter to constrain. To obtain an
estimate of the lower bound, one may use the method, first suggested
by Bahcall (1967), whereby one places an upper limit on $n_{p}$, the
number density of protons in the cloud, using the equation,
\begin{equation}
\label{c2}
\frac{N({\rm C}{\sc \,ii}^*)}{N({\rm C}{\sc \,ii})} = 3.9 \times
10^{-2}n_e\left(1+0.22\frac{n_p}{n_e}\right).
\end{equation}
Here, $N({\rm C}{\sc \,ii}^*)$ is the column density of C{\sc \,ii} which
has been collisionally excited to the state C{\sc \,ii}$^*$. $n_e$ is the
electron number density. We assume that $n_e \sim n_p$ and that $n_p \sim
n_{\rm H}$, and so, using $N({\rm C}{\sc \,ii}) = 4.6 \times 10^{14}{\rm
~cm}^{-2}$ and $N({\rm C}{\sc \,ii}^*) < 10^{12}{\rm ~cm}^{-2}$ from
Songaila et al. (1994), we estimate that $\log_{10}n_{\rm H} < -1.3$. Under
these simple assumptions, this value is consistent with Morris et
al. (1986) estimate, $\log_{10}n_{\rm H} < 0.0$, and we adopt this as a
conservative upper limit. A lower limit may be obtained by using upper
limits on the cloud sizes. We used results from Bergeron \& Stasi\'{n}ska
(1986) to place a lower limit of $\log_{10}n_{\rm H} > -3$ where we have
used {\sc cloudy} to estimate that $N({\rm H}{\sc \,i}) \sim N({\rm
H})/1000$.  We compare our parameter range estimates above with those
assumed in Lopez et al. (1999) and find that they are consistent after
considering the differences between the assumed radiation field.

We present the results of the grid of {\sc cloudy} models in Table 1
where we only give the value of $\Gamma_{X,{\rm Mg}{\sc \,ii}}^{\rm
max}$, the maximum value of $\Gamma_{X,{\rm Mg}{\sc \,ii}}$ that
occurred in the grid. We also present the model responsible for this
value of $\Gamma_{X,{\rm Mg}{\sc \,ii}}$. All values of
$\Gamma_{X,{\rm Mg}{\sc ii}}^{\rm max}$ should be treated as having an
error of at least 0.3 dex. We find that these results are consistent
with the predictions of {\sc model} 1 in Lopez et al. (1999) and are
also consistent with their measured values.

\begin{table*}
\caption{Results of the {\sc cloudy} photo-ionization equilibrium
calculation showing the upper limit on $\Gamma_{X,{\rm Mg}{\sc \,ii}} =
\log_{10}\frac{N(X)}{N({\rm Mg}{\sc ii})}$. The model used is specified by
the logarithmic parameters, $U$, $N$(H{\sc \,i}), $n_{\rm H}$ and $Z$.}
\label{tab:cloudy}
\begin{tabular}{lccccrlccccr} \hline
Ion    &   $\Gamma_{X,{\rm Mg}{\sc \,ii}}^{\rm max}$ & $U$  &  $N({\rm H}{\sc \,i})$  & $n_{\rm H}$ &  $Z$ & Ion    &   $\Gamma_{X,{\rm Mg}{\sc \,ii}}^{\rm max}$ & $U$  &  $N({\rm H}{\sc \,i})$  & $n_{\rm H}$ &  $Z$ \\ \hline 
Fe{\sc \,i}  	& -0.20 &  -4 &   19 &    -3  &   -2&  Ni{\sc \,i}     & -2.39 &  -4 &   17 &    -3  &   -2   \\
Fe{\sc \,ii}  	&  0.62 &  -3 &   19 &    -3  &    0& 	Ni{\sc \,ii}    & -0.79 &  -3 &   19 &    -3  &    0   \\
Fe{\sc \,iii} 	&  0.84 &  -3 &   17 &    -3  &   -2& 	Ni{\sc \,iii}   & -1.07 &  -3 &   17 &    -3  &   -2   \\
Fe{\sc \,iv}  	&  0.99 &  -3 &   17 &    -3  &   -2&  Ni{\sc \,iv}    & -0.63 &  -3 &   17 &     0  &   -2   \\
Fe{\sc \,v}  	&  0.42 &  -3 &   17 &    -3  &   -2& 	Ni{\sc \,v}     & -1.05 &  -3 &   17 &    -3  &   -2   \\
Fe{\sc \,vi}  	&  0.11 &  -3 &   17 &    -3  &   -2& 	Ni{\sc \,vi}    & -0.62 &  -3 &   17 &    -3  &   -2   \\
Fe{\sc \,vii} 	& -0.71 &  -3 &   17 &    -3  &   -2& 	Ni{\sc \,vii}   & -0.92 &  -3 &   17 &    -3  &   -2   \\
Fe{\sc \,viii}	& -2.07 &  -3 &   17 &    -3  &   -2& 	Ni{\sc \,viii}  & -1.79 &  -3 &   17 &    -3  &   -2   \\
 & & & & & & & & & & &	               	       	     \\  	     
Mg{\sc \,i}   	& -1.50 &  -4 &   19 &    -3  &   -2&   Na{\sc \,i}     & -2.18 &  -4 &   17 &    -3  &    0   \\
Mg{\sc \,iii} 	&  1.16 &  -3 &   17 &    -3  &   -2& 	 Na{\sc \,ii}    & -0.80 &  -3 &   17 &     0  &   -2   \\
Mg{\sc \,iv}  	&  0.19 &  -3 &   17 &    -3  &   -2& 	 Na{\sc \,iii}   & -0.23 &  -3 &   17 &    -3  &   -2   \\
Mg{\sc \,v}  	& -1.24 &  -3 &   17 &    -3  &   -2& 	 Na{\sc \,iv}    & -0.74 &  -3 &   17 &    -3  &   -2   \\
Mg{\sc \,vi} 	& -2.20 &  -3 &   17 &    -3  &   -2& 	 Na{\sc \,v}     & -1.82 &  -3 &   17 &    -3  &   -2   \\
 & & & & & & & & & & &	               	       	     \\  	     
Al{\sc \,i}  	& -2.02 &  -3 &   17 &    -3  &   -2&   Si{\sc \,i}     & -1.53 &  -4 &   17 &    -3  &    0   \\
Al{\sc \,ii}   	& -0.17 &  -3 &   17 &    -3  &   -2& 	 Si{\sc \,ii}    &  0.62 &  -3 &   17 &    -3  &    0   \\
Al{\sc \,iii} 	& -0.64 &  -3 &   17 &     0  &    0& 	 Si{\sc \,iii}   &  1.09 &  -3 &   17 &    -3  &   -2   \\
Al{\sc \,iv}  	& -0.31 &  -3 &   17 &     0  &    0& 	 Si{\sc \,iv}    &  0.26 &  -3 &   17 &     0  &   -2   \\
Al{\sc \,v}  	& -1.37 &  -3 &   17 &     0  &    0& 	 Si{\sc \,v}     & -0.62 &  -3 &   17 &     0  &   -2   \\
Al{\sc \,vi}  	& -2.28 &  -3 &   17 &     0  &    0& 	 Si{\sc \,vi}    & -1.46 &  -3 &   17 &     0  &   -2   \\
 & & & & & & & & & & &	               	       	     \\  	     
Ca{\sc \,i}   	& -2.03 &  -4 &   17 &    -3  &    0&   C{\sc \,i}      & -0.33 &  -4 &   17 &    -3  &   -2   \\
Ca{\sc \,ii}  	& -1.32 &  -4 &   17 &     0  &    0& 	 C{\sc \,ii}     &  1.53 &  -3 &   17 &    -3  &   -2   \\
Ca{\sc \,iii} 	& -0.01 &  -3 &   17 &    -3  &   -2& 	 C{\sc \,iii}    &  2.08 &  -3 &   17 &    -3  &   -2   \\
Ca{\sc \,iv}  	& -0.84 &  -3 &   17 &    -3  &   -2& 	 C{\sc \,iv}     &  0.85 &  -3 &   17 &    -3  &   -2   \\
Ca{\sc \,v}   	& -2.18 &  -3 &   17 &    -3  &   -2& 	 C{\sc \,v}      &  0.00 &  -3 &   17 &    -3  &   -2   \\
 & & & & & & & & & & &	               	       	     \\  	     
N{\sc \,i}    	&  0.70 &  -4 &   19 &     0  &    0&   O{\sc \,i}      &  1.79 &  -3 &   19 &    -3  &    0   \\
N{\sc \,ii}   	&  1.01 &  -3 &   17 &    -3  &   -2& 	 O{\sc \,ii}     &  1.88 &  -3 &   19 &    -3  &    0   \\
N{\sc \,iii}	&  1.57 &  -3 &   17 &    -3  &   -2& 	 O{\sc \,iii}    &  2.47 &  -3 &   19 &    -3  &    0   \\
N{\sc \,iv}  	&  0.48 &  -3 &   17 &    -3  &   -2& 	 O{\sc \,iv}     &  1.13 &  -3 &   19 &    -3  &    0   \\
N{\sc \,v}    	& -1.03 &  -3 &   17 &    -3  &   -2& 	 O{\sc \,v}      & -0.21 &  -3 &   19 &    -3  &    0   \\
N{\sc \,vi}   	& -1.89 &  -3 &   17 &    -3  &   -2& 	 O{\sc \,vi}     & -1.70 &  -3 &   19 &    -3  &    0   \\
 & & & & & & & & & & &	               	       	     \\  	     
Ne{\sc \,i}   	& -0.22 &  -4 &   19 &    -3  &    0&   Ar{\sc \,i}     & -1.24 &  -4 &   19 &    -3  &   -2   \\
Ne{\sc \,ii}  	&  0.57 &  -3 &   19 &    -3  &    0& 	 Ar{\sc \,ii}    & -0.53 &  -3 &   17 &    -3  &   -2   \\
Ne{\sc \,iii} 	&  1.68 &  -3 &   17 &    -3  &   -2& 	 Ar{\sc \,iii}   &  0.01 &  -3 &   17 &     0  &   -2   \\
Ne{\sc \,iv}  	&  0.68 &  -3 &   17 &    -3  &   -2& 	 Ar{\sc \,iv}    & -0.89 &  -3 &   17 &     0  &   -2   \\
Ne{\sc \,v}   	& -0.70 &  -3 &   17 &    -3  &   -2& 	 Ar{\sc \,v}     & -2.38 &  -3 &   17 &    -3  &   -2   \\ \hline
\end{tabular}
\end{table*}

\subsubsection{A Search for Interlopers and Results}

Having obtained estimates of the column density upper limits (Table
1), we can use the results of Fig. 3 to restrict the possible values
for $\Delta\lambda$ for any of the species above.  Fig. 3 shows the
value of this test, since, given a value for $\Gamma_{X,{\rm Mg}{\sc
\,ii}}$, we can place very stringent constraints on the allowed range
in $\Delta\lambda$.  Then, we search tabulated transitions for each
species to see if any candidate interlopers exist.

Table 1 and the results expressed in Fig. 3 show that any interloper must
have a wavelength differing by no more than $\sim 0.1$\AA\, to that of
Mg{\sc \,ii} $\lambda2796$. We searched published energy level tables
(Moore 1971; the Vienna Atomic Line Database (VALD) -- Piskunov et al. 1995
and Kupka et al. 1999) to find potential lines of those species shown in
Table \ref{tab:cloudy} which satisfy the above criteria.  We searched only
for E1 transitions as all other transitions are associated with a large
factor of suppression. For example, E2 and M1 transition probabilities are
suppressed by factors $\sim \alpha^{-2}$ and $\sim
\alpha^{-2}(Z_n\alpha)^{-4}$ respectively ($Z_n$ is the nuclear
charge). First-forbidden E1 transitions can also be considered since the
oscillator strengths are typically $\sim 10Z_n^4\alpha^4$ times that of a
normal E1 transition in a related or similar ion.

Oscillator strengths, $f_X$, are not generally available for the
potentially interloping transitions, so we must adopt an upper limit
by using other (stronger) transitions for the same multiplet (for
which oscillator strengths are known).

Given upper limits on $f_X$ and $\Gamma_{X,{\rm Mg}{\sc \,ii}}$ and a
value for $\Delta\lambda$, we can estimate $d\lambda$ and hence derive
a conservative upper limit on the interloper's effect on $\da$.

Although Fig. 3 relates specifically to Mg{\sc \,ii} $\lambda2796$, similar
simulations were carried out to search for interlopers near all other
`host' absorption lines (i.e. those in table 1 in M01a).  The model
parameters were also extended to DLAs (e.g. $19 < \log_{10}N$(H{\sc \,i})$
< 22$).

We have {\it not been able to identify any candidate interlopers}
satisfying all of the criteria above.

We note that {\sc cloudy} only computes column densities for the most
abundant species.  However, it is possible that other low abundance species
could mimic a shift in $\alpha$. Therefore, we searched for interloping
lines (in Moore 1971 and VALD) of ionic species of elements such as Sc, Ti,
V, Cr, Mn, Co and Zn and found only one possibility: a Cr{\sc \,ii}
transition with $\lambda_0 = 2026.2686{\rm ~\AA}$ which lies $0.13{\rm
~\AA}$ to the red of the Zn{\sc \,ii} $\lambda 2026$ `host' line. This line
has an oscillator strength of $0.047$ according to VALD.  We simulated its
effect on $\da$ for the \hiz~sample.

We obtained the largest value of $N({\rm Cr}\,{\sc \,ii})/N({\rm Zn}\,{\sc
\,ii})$ directly from the observed QSO absorption line data.  We then
generated a synthetic, single velocity component spectrum containing Zn{\sc
\,ii} $\lambda2026$ and the Cr{\sc \,ii} interloper.  We also generated
several other lines seen in the \hiz~sample. We produce synthetic spectra
with several different values of $\da$ and then used {\sc vpfit} to measure
$\da$. The result was that the measured value of $\da$ was always
consistent with the nominal value, that is, we could detect no effect due
to the interloping Cr{\sc \,ii} transition.

In summary, we have found no interloping species which could have caused a
significant shift in $\da$. We do note, however, that the ionic energy
level data in Moore (1971) and VALD are incomplete for many highly ionized
species and so we cannot rule out the possibility of blending with
undiscovered transitions. Also, as noted previously, this analysis only
takes into account blends with atomic, not molecular species. Firstly,
typical oscillator strengths for molecular transitions are much smaller
than for resonance atomic transitions. Also, the typical column densities
of the low abundance elements in Table \ref{tab:cloudy} suggest that very
few molecules will exist in the QSO absorption clouds. Therefore, we
consider the possible effect of any molecular blending to be negligible.

\subsection{Line removal as a test for systematic blends}\label{sec:linrem}
If a particular host transition (e.g. Mg{\sc \,ii} $\lambda$2796) were
systematically blended then removing that transition from the analysis in
M01a and W01 should result in a different value of $\da$. Line removal is
also a rough test for isotopic ratio or hyperfine structure effects as
introduced in Sections \ref{sec:sat}--\ref{sec:hyper}.  We therefore remove
a particular transition in an absorption complex only if doing so still
allows us to obtain meaningful constraints on $\da$. For example, we do not
remove the Mg{\sc \,ii} $\lambda$2796 line from a fit to a Mg{\sc
\,ii}/Fe{\sc \,ii} system if the Mg{\sc \,ii} $\lambda$2803 line is not
present. We also removed entire species where possible.  A small difficulty
occurs with Cr{\sc \,ii} and Zn{\sc \,ii} $\lambda$2062, whose profiles are
sometimes blended together (i.e. when the velocity structure is broad). If
the profiles were blended then we simply removed both transitions
simultaneously, only removing them separately when the lines were
completely separated.  We obtain $\da$ with the same procedure described in
M01a.

Removing a transition from the fit may result in a slightly modified
estimate for the velocity structure, revising the estimate for $\da$ for
that system.  The question is, how robust is the estimate for $\da$,
averaged over the entire sample, to this line removal process?

We present our results in Fig. 4. For each transition removed, we
compare the weighted mean of $\da$ (and the associated 1$\sigma$
error) before and after the line removal. It must be stressed that
both the value and the errors are not independent and so interpreting
Fig. 4 is difficult. However, it seems that there are no extreme
deviations after line removal. Thus, we might conclude from such a
diagram that systematic blending has not occurred and that we have no
reason to suspect that one or more of the laboratory wavelengths is
incorrect.

Concentrating on the heavier ions, we might also conclude that there is no
effect on $\da$ due to possible isotopic ratio or hyperfine structure
effects. We cannot draw similar conclusions for Mg since we cannot remove
all Mg lines simultaneously (this would prevent us from obtaining
meaningful constraints on $\da$ for the \loz~sample). We also see that any
lack of thermal equilibrium in the absorption clouds, leading to unequal
population of the hyperfine levels of Al{\sc \,iii} (see Section
\ref{sec:hyper}), has had no significant effect on $\da$. The large error
bars on the Al{\sc \,iii} points reflect the fact that those lines are only
present in 6 of our \hiz~systems.

\begin{figure}
\centerline{\psfig{file=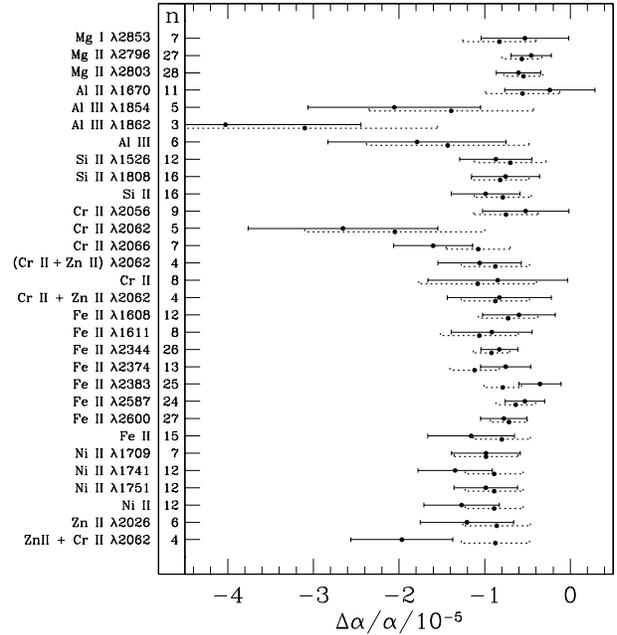,width=3.5in}}
\caption{Comparison of the weighted mean of $\da$ before (dotted error bar)
and after (solid error bar) line removal. The transitions removed are
listed on the left together with the number of systems, $n$, for which
removal of that transition was possible. It is stressed that comparing the
values and error bars before and after line removal is difficult since
these quantities are not independent. Note that there may be some confusion
due to the occasional blending of the Cr{\sc \,ii} and Zn{\sc \,ii}
$\lambda$2062 lines: `(Cr{\sc \,ii} + Zn{\sc \,ii}) $\lambda$2062' refers
to cases where both transitions had to be removed simultaneously; `Cr{\sc
\,ii} + Zn{\sc \,ii} $\lambda2062$' refers to cases when all Cr{\sc \,ii}
transitions were removed along with the blended Zn{\sc \,ii} line; a
similar definition applies to `Zn{\sc \,ii} + Cr{\sc \,ii} $\lambda2062$';
`Cr{\sc \,ii}' then refers only to the removal of all Cr{\sc \,ii}
transitions in cases where there was no blending with the Zn{\sc \,ii}
line. Only one similar case occurred for removal of all Zn{\sc \,ii} lines
and so we do not present this result.}
\end{figure}

Qualitatively, this line removal test suggests that there is no one
single `host' transition or species which is systematically blended
and which has therefore affected our results in any significant way.
The results also suggest that isotopic ratio or hyperfine structure
effects have not significantly effected the heavier ions in our
analysis. This point is more fully explored in Section
\ref{sec:saturation} and \ref{sec:noiso}.

\section{Detailed analysis: atmospheric dispersion effects}\label{sec:atmoref}
\begin{figure}
\centerline{\psfig{file=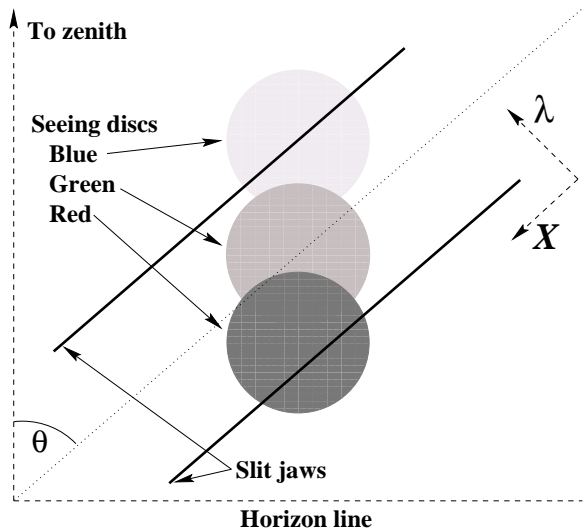,width=3in}}
\caption{Schematic diagram showing the effects of atmospheric
dispersion. The HIRES slit is projected onto the sky so that we see the
seeing discs of blue, red and green light from a point source
dispersed with some component along the spectral direction of the slit. The
spectral direction is indicated by $\lambda$ and the spatial direction by
$X$. Note that the refraction is worse in the blue.}
\end{figure}

The problems that arise due to atmospheric dispersion effects can be
understood with reference to Fig. 5. If we observe a point source with
the spectrograph slit at some angle $\theta$ to the vertical (as
projected on the sky) then the object is dispersed with some component
along the spectral direction of the slit. The optical design of the
Keck/HIRES combination at the time at which a large fraction of our
data was taken was such that $\theta = \xi$ for $\xi$ the zenith angle
of the object (Tom Bida, Steve Vogt, personal communication). As we
noted in Section \ref{sec:atmodisp}, this leads to two effects on the
wavelength scale:

\begin{enumerate}
\item The first effect is a stretching of the spectrum (relative to the
ThAr calibration frames) due to the angular separation of different
wavelengths entering the slit. Consider two wavelengths, $\lambda_1$ and
$\lambda_2$ ($\lambda_2 > \lambda_1$), falling across the slit as in
Fig. 5. If we were to measure the spectral separation between these two
wavelengths on the CCD, $\Delta\lambda'$, we would find that
\begin{equation}\label{eq:angsep}
\Delta\lambda' \approx \lambda_2-\lambda_1 +\frac{a\Delta\psi
\sin{\theta}}{\delta}\, ,
\end{equation}
where $a$ is the CCD pixel size in \AA ngstroms, $\delta$ is the projected
slit width in arc seconds per pixel (for HIRES, $\delta = 0\farcs287$ per
pixel) and $\Delta\psi$ is the angular separation (in arc seconds) of
$\lambda_1$ and $\lambda_2$ at the slit. $\Delta\psi$ is a function of the
atmospheric conditions along the line of sight to the object and can be
approximated using the refractive index of air at the observer and the
zenith distance of the object (e.g. Filippenko 1982). Note that equation
\ref{eq:angsep} assumes that the seeing profile is not truncated at the
slit edges.  If tracking errors or seeing effects cause truncation,
equation \ref{eq:angsep} provides an upper limit to the measured
$\Delta\lambda'$.

\item If truncation of the seeing profile does occur, it may be
asymmetric (as the diagram illustrates).  In the case of HIRES, the
optical design is such that a blue spectral feature will have its red
wing truncated and a red spectral feature will be truncated towards
the blue. This is the case shown in Fig. 5. Thus, an asymmetry is
introduced into the IP and this will be wavelength dependent. The
extent of the asymmetry will depend on $\Delta\psi$ (which depends
most strongly on $\xi$).
\end{enumerate}

Note that when we centroid spectral features to find their separation, both
effect (i) (shifting) and effect (ii) (instrumental profile (IP)
distortion) effectively `stretch' the spectrum.  For the \loz~sample, this
should lead to systematically positive values of $\da$. That is,
atmospheric dispersion effects cannot explain the average negative value of
$\da$ seen in the QSO data. Removing these effects from our data would make
the results {\it more} significant. Although there is no obvious
correlation between $\da$ and $\xi$ as would be expected in this scenario
(see Fig. 6), the potential effect is quantified in our analysis below.

\begin{figure}
\centerline{\psfig{file=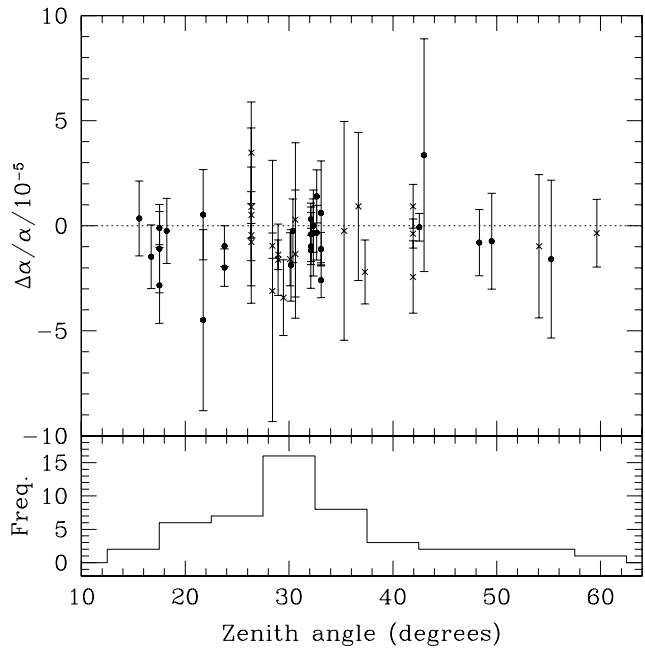,width=3.5in}}
\caption{The lower panel shows the distribution of absorption systems over
average zenith distance. The upper panel shows the QSO data as a
function of zenith distance. We do not see any obvious correlation. When we
remove points for which there should be no atmospheric dispersion effect,
the Spearman rank correlation coefficient is 0.18 with an associated
probability of 0.41. The coefficients for the low (filled circles) and high
(crosses) redshift samples are 0.16 and -0.09 with probabilities 0.43 and
0.71 respectively.}
\end{figure}

For all objects which were observed without the use of an image rotator, we
have used the observational parameters reported in the image header files
to find average values of $\xi$ (see Fig. 6), temperature, pressure and
relative humidity for each object. This allows us to correct for the
stretching of the spectrum implied by equation \ref{eq:angsep}. We have
re-calculated $\da$ in the same manner as described in M01a and we plot the
results in Fig. 7.  For several QSOs, only a single image header (i.e. for
one of the QSO exposures out of a series) was available. In these cases we
have added (in quadrature) an error of $0.9 \times 10^{-5}$.  This is an
average value determined by calculating $\da$ for several values of the
zenith distance for those objects concerned and corresponds to an
uncertainty in the average $\xi$ of about $\pm 20^{\circ}$.

We have also tested the effect of the IP distortions (effect (ii) above) on
$\da$. We generated high SNR, single component absorption spectra (with
$\da = 0$) for several redshifts. We included different combinations of
transitions in the different spectra so as to make a representative
estimate for both samples of QSO data (particularly the \hiz~sample which
contains many combinations of different transitions). A wavelength
dependent PSF was constructed using the following procedure. We assume that
the seeing (taken as 0\farcs8) and tracking error (taken as 0\farcs25)
generate Gaussian profiles across the slit and that a wavelength
$\lambda_0$ falls at its centre.  Other wavelengths are refracted to
different parts of the slit depending on the zenith distance.  To compute
this, typical Keck atmospheric conditions (temperature, pressure, relative
humidity) were used.

We truncate the profile by multiplying by a top-hat function (whose width
is equal to that of the slit) and then convolve the result with a Gaussian
IP of width $\sigma = 2.2{\rm ~kms}^{-1}$ (Valenti et al. 1995). This final
PSF is then convolved with the synthetic absorption spectra. The value of
$\lambda_0$ is $\sim 5500{\rm ~\AA}$ (i.e. the TV acquisition system is
centred at about this wavelength).  Using this value for $\lambda_0$ and
$\xi = 45^{\circ}$ we find $\da$ to be quite insensitive to the IP
distortions, particularly in the \hiz~regime. The effect on $\da$ may be as
large as $0.05 \times 10^{-5}$ for the \loz~regime but round-off errors in
{\sc vpfit} become important at this level of precision. To exaggerate the
effect, we also chose extreme values of $\lambda_0$ to be 4000 and
7000\AA. The effect on $\da$ in these cases was still negligible in the
\hiz~regime but increased to the level of $\da \sim 0.12 \times 10^{-5}$
for the \loz~spectra.  This latter value corresponds to $\sim$1/6 of the
correction due to the shifting effect.  The sign of the above corrections
to $\da$ are such that applying them to the QSO data make the observed
average $\da$ {\it more negative}.

In summary, we find $\da$ to be most sensitive to the shifting effect
described by equation \ref{eq:angsep} and reasonably insensitive to the IP
distortion effect. Thus, we do not apply a correction for the latter in
the results shown in Fig. 7. The results show the values of $\da$ once
the shifting effect has been removed from the data. As expected, we
see a significant decrease in $\da$ in the \loz~sample. The
\hiz~sample is insensitive to this systematic effect, as can
be expected due to the many different transitions used with $q_1$
coefficients of different signs. We summarise the results in Table
\ref{tab:da} and compare them with those found in M01a and W01.  We stress
that the corrections above (equation \ref{eq:angsep}) are upper limits.
Seeing and tracking errors will diminish the effects substantially.

\begin{figure}
\centerline{\psfig{file=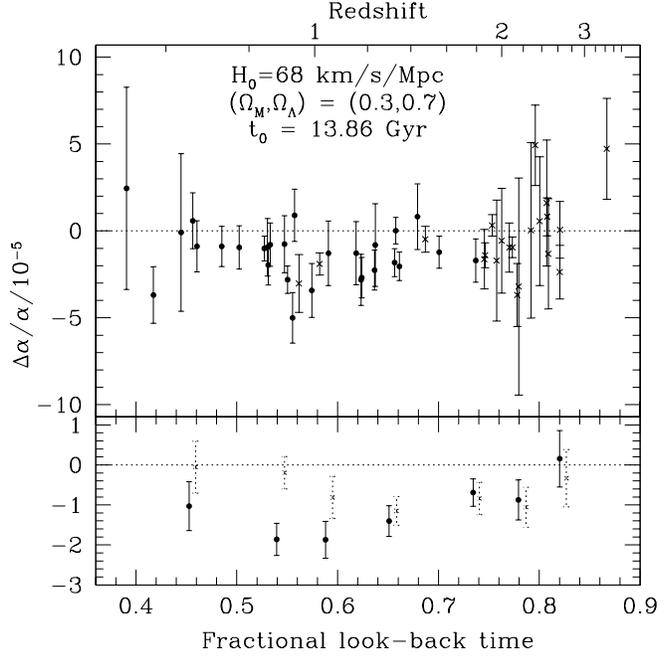,width=3.5in}}
\caption{Results after removal of atmospheric dispersion effects. The top
panel shows our raw results for $\da$ as a function of look-back time in a
flat, $\Lambda$ cosmology. The redshift scale is also provided for
comparison. The lower panel shows an arbitrary binning of the results which
emphasizes the susceptibility of the \loz~sample to this systematic
error (the dotted lines are the results of M01a and W01, slightly shifted for
clarity). Note, however, that the correction made here is an extreme and
will be diminished by seeing and tracking effects.}
\end{figure}

\begin{table}
\caption{Comparison of the weighted means, $\left<\da\right>_{\rm w}$,
with the results of M01a and W01, after correction for atmospheric dispersion
effects.}
\begin{center}
\label{tab:da}
\begin{tabular}{lcc}\hline
Sample    & $\left<\da\right>_{\rm w}/10^{-5}$ & $\left<\da\right>^{\rm M01a}_{\rm w}/10^{-5}$\\\hline
Low-$z$   & $-1.52\pm 0.23$ & $-0.70\pm 0.23$\\
High-$z$  & $-0.79\pm 0.26$ & $-0.76\pm 0.28$\\
Total     & $-1.19\pm 0.17$ & $-0.72\pm 0.18$\\\hline
\end{tabular}
\end{center}                            
\end{table}

\section{Detailed analysis: differential isotopic saturation}\label{sec:saturation}
In this section we quantify the effect of differential isotopic saturation
on our measured values of $\da$. As discussed in Section \ref{sec:sat},
this comes about when the strongest isotopic component of an absorption
line saturates. The weaker isotopic components may still be on the linear
part of their curves of growth.  Using a single weighted wavelength to
represent the composite profile position (i.e. the profile centroid) can
therefore, in principle, result in systematic errors in $\da$.

The various isotopic abundances of concern are shown in Table
\ref{tab:iso}. If only composite wavelengths were used then, if there
exists a saturated dominant isotope in the QSO data, this may result in
systematic errors in $\da$. However, composite wavelengths were only used
for the 4 heaviest atoms (i.e. Cr, Fe, Ni, Zn) and, of these, only some Fe
transitions appear to be saturated in some of our QSO spectra.  The Cr, Ni
and Zn lines were always very weak (typically absorbing only $\sim$20\% of
the continuum) and so using composite transition wavelengths is justified.

\begin{table}
\caption{The percentage terrestrial abundances of the atoms used in our
analysis (Rosman \& Taylor 1998). The neutron number is defined relative to
that of the isotope with the largest abundance and is denoted as $\Delta A$
(negative values representing lighter isotopes).}
\begin{center}
\label{tab:iso}
\begin{tabular}{lccccccc}\hline
Atom    & \multicolumn{7}{c}{$\Delta A$}\\
        &-2   &0     &+1   &+2   &+3  &+4  &+6  \\\hline
Mg      &     &78.99 &10.00&11.01&    &    &    \\
Al      &     &100.0 &     &     &    &    &    \\
Si      &     &92.23 &4.68 &3.09 &    &    &    \\
Cr      &4.345&83.789&9.581&2.365&    &    &    \\
Fe      &5.845&91.754&2.119&0.282&    &    &    \\
Ni      &     &68.08 &     &26.22&1.14&3.64&0.93\\ 
Zn      &     &48.6  &     &27.9 &4.1 &18.8&0.6 \\\hline
\end{tabular}
\end{center}                            
\end{table}

It is therefore left to quantify the effect of differential isotopic
saturation for the Fe lines in our spectra. Ideally, if we knew the
isotopic wavelength separations for the various Fe{\sc \,ii}
transitions then we could do this by simply fitting our spectra with
this structure to find $\da$. The difference between the weighted mean
$\da$ and that found in M01a and W01 would indicate the size of the effect. In
reality, we do not know the isotopic separations. We can, however,
place an upper limit on this effect since we know (or can estimate)
the isotopic separations of the Mg and Si anchor lines. If we fit our
spectra with only the composite Mg and Si wavelengths then we will
obtain maximal changes to $\da$ since (i) these atoms have strong
lines which saturate more often than the Fe{\sc \,ii} lines and (ii)
they will have the largest isotopic separations (i.e. largest mass
shift).

We have carried out the above re-calculation of $\da$ using the composite
wavelengths for the Mg{\sc \,i}, Mg{\sc \,ii} and Si{\sc \,ii} lines. In
the \loz~sample, the fact that the Mg lines have their weaker isotopes at
shorter wavelengths means that $\da$ should be systematically more negative
if the composite wavelengths are used. The situation is again more
complicated in the \hiz~sample. We find only very small differences between
the results and those of M01a and W01, the weighted means for the low and
\hiz~sample being $\da = (-0.75 \pm 0.23) \times 10^{-5}$ and $\da = (-0.74
\pm 0.28) \times 10^{-5}$ respectively (compared to $\da = (-0.70 \pm 0.23)
\times 10^{-5}$ and $\da = (-0.76 \pm 0.28) \times 10^{-5}$ from M01a and
W01).

Can we understand the magnitude of these corrections? Imagine a single
velocity component and consider finding $\da$ from, say, the Mg{\sc
\,ii} $\lambda$2796 and Fe{\sc \,ii} $\lambda$2344 lines only. One
might estimate the maximum effect of differential isotopic saturation
by assuming that when all isotopes of Mg{\sc \,ii} are saturated, the
line centroid would lie at the unweighted mean isotopic
wavelength. This would result in a correction to $\da$ of $\sim -1
\times 10^{-5}$ (i.e. the observed value would become more negative
when the correction is applied). 

However, several factors will reduce this correction in reality.  Firstly,
once a line begins to saturate, then because the intrinsic line width
(i.e. the $b$-parameter) is far greater than the isotopic separations, the
weaker isotopes are swamped by the dominant one. This reduces the effect on
$\da$.  Secondly, if the velocity structure is more complex (i.e. if it
contains more than one velocity component) the unsaturated components,
which do not suffer from the differential isotopic saturation problem, will
provide much stronger constraints on $\da$.  Finally, as more unsaturated
transitions (e.g. Fe{\sc \,ii}) are incorporated into the fit of a
particular QSO absorption system, the effect is further reduced since some
constraints on $\da$ come from shifts between the Fe{\sc \,ii} lines.

It is therefore clear that differential saturation of the isotopic
components of the various Fe{\sc \,ii} lines cannot have lead to any
significant effect on $\da$. Our results also indicate that the weighted
mean $\da$ for the \hiz~sample will not be very sensitive to
errors in our estimate of the Si{\sc \,ii} isotopic separations.

\section{Detailed analysis: isotopic abundance variation}\label{sec:noiso}
The analysis we have carried out so far assumes terrestrial isotopic
abundances.  However, it is quite possible, even likely, that these high
redshift gas clouds have rather different abundances. As we noted in
Section \ref{sec:iso}, a substantial systematic shift in $\alpha$ might be
mimicked and will be particularly pronounced in the \loz~sample
since the Mg isotopic separations are the largest. No observations of high
redshift isotopic abundances have been made so we have no {\it a priori}
information on the QSO absorption systems themselves.  However,
observations of Mg (Gay \& Lambert 2000) and theoretical estimates for Si
(Timmes \& Clayton 1996) in stars clearly show a decrease in isotopic
abundances with decreasing metallicity.  For example, at relative metal
abundances [Fe/H] $\sim -1$, $^{25}$Mg/$^{24}$Mg $\sim$ $^{26}$Mg/$^{24}$Mg
$\approx 0.1$, i.e. about 20\% below terrestrial values (see Table
\ref{tab:iso}). Theoretical calculations (Timmes, Woosley \& Weaver 1995)
suggest even larger decreases.

Estimates have also been made of the metal abundances in a subset of the
QSO absorption systems used in our analysis. The DLAs have a mean
metallicity of $Z \sim -1.5$ (Prochaska \& Wolfe 2000) and the Mg/Fe
systems span the range $Z=-2.5$ to 0.0 (Churchill et al. 1999).

Therefore, we explored the effect of isotopic abundance evolution on $\da$
for the entire range in isotopic abundance from zero to terrestrial values
(i.e. in the case of Mg, the limiting case is $^{25}$Mg/$^{24}$Mg =
$^{26}$Mg/$^{24}$Mg = 0). This allows us to derive a secure upper limit on
the magnitude of the effect on $\da$.

We have re-computed $\da$ for the entire sample after removing the
following isotopes: $^{25}$Mg, $^{26}$Mg, $^{29}$Si and $^{30}$Si. The
wavelengths for the dominant isotopic components, $^{24}$Mg and $^{28}$Si,
for the various transitions of interest, are given in table 3 of M01a and
are summarized in Table \ref{tab:isowavs}.

\begin{table}
\begin{center}
\caption{Wavelengths of the dominant isotopic components of the Mg and Si
transitions. Those of Mg are derived from experimental data in Pickering et
al. (1998) while those of Si are calculated in M01a on the basis of a
scaling by the mass shift from the basic Mg isotopic structure.}
\label{tab:isowavs}
\begin{tabular}{lccc}\hline
Transition                 & $m$ & $\omega_0$ & $\lambda_0$\\\hline
Mg{\sc \,i} $\lambda$2853  & 24  & 35051.271  & 2852.9636  \\
Mg{\sc \,ii} $\lambda$2803 & 24  & 35669.286  & 2803.5324  \\
Mg{\sc \,ii} $\lambda$2796 & 24  & 35760.835  & 2796.3553  \\
Si{\sc \,ii} $\lambda$1808 & 28  & 55309.330  & 1808.0132  \\
Si{\sc \,ii} $\lambda$1526 & 28  & 65500.442  & 1526.7073  \\\hline
\end{tabular}
\end{center}
\end{table}

The results are presented in Table \ref{tab:noiso} and Fig. 8. These show,
as expected, that the \loz~sample is more sensitive to this systematic
error than the \hiz~sample. We stress that the signs of the above
corrections are such that applying them to the QSO data make the observed
average $\da$ {\it more negative}. Since we have no information on the
isotopic separations for the other atoms of interest, we cannot perform a
similar analysis for all species. However, the corrections due to variation
in the isotopic abundances of Mg and Si should be dominant (as we discuss
is Section \ref{sec:iso}) so the corrections shown in Table \ref{tab:noiso}
and Fig. 8 are conservative upper limits.

\begin{table}
\caption{Comparison of the weighted means, $\left<\da\right>_{\rm w}$, with
the results of M01a and W01, after maximal correction for possible variation in
isotopic ratios}
\begin{center}
\label{tab:noiso}
\begin{tabular}{lcc}\hline
Sample    & $\left<\da\right>_{\rm w}/10^{-5}$ & $\left<\da\right>^{\rm M01a}_{\rm w}/10^{-5}$\\\hline
Low $z$   & $-1.01\pm 0.23$ & $-0.70\pm 0.23$\\
High $z$  & $-0.87\pm 0.26$ & $-0.76\pm 0.28$\\
Total     & $-0.96\pm 0.17$ & $-0.72\pm 0.18$\\\hline
\end{tabular}
\end{center}                            
\end{table}

\begin{figure}
\label{fig:noiso}
\centerline{\psfig{file=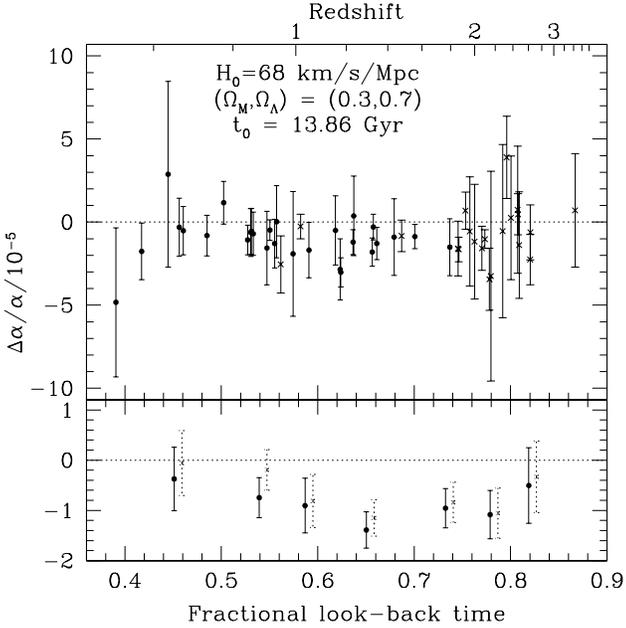,width=8.5cm}}
\caption{Results after removing the weaker isotopic components of Mg and
Si. The upper panel shows our raw results and the lower panel shows an
arbitrary binning which emphasizes the susceptibility of the \loz~sample to
this systematic error (the dotted lines are the results of M01a and W01,
slightly shifted for clarity). Note, however, that the correction made here
is extreme.}
\end{figure}

\section{A general test for systematic errors}\label{sec:apns}

Although we have already convincingly ruled out any significant
wavelength-scale distortion using ThAr spectra (Section
\ref{sec:wavcal}), the QSO and ThAr light does not follow precisely
the same light path.  Indeed, for these reasons we had to consider
some of the previous effects such as atmospheric dispersion.  Thus, it
is conceivable that some residual long-scale wavelength distortion or
other simple systematic error remains in the QSO spectra. 

The dependence of $\da$ on the $q_1$ coefficients is simple for the
\loz~sample but considerably more complicated for the high z
sample.  This can be seen by inspecting table 1 in M01a.  If
systematic errors are responsible for the non-zero $\da$ we measure,
it is therefore surprising that $\da$ is so consistent between the 2
samples.  {\it A priori}, it would be even more surprising if we were
to take subsets of the QSO data, grouped according to the sign and
magnitude of $q_1$, and found a {\it consistent} value for $\da$.

We apply this idea to the \hiz~sample only, since it contains transitions
with a wide range of $q_1$ values.  Of the 21 absorbers in this sample, we
can find a subset of 13 systems that include at least one anchor line
(transitions with small $q_1$), at least one positive-shifter ($q_1 \ge
700{\rm ~cm}^{-1}$) {\it and} at least one negative-shifter ($q_1 \le
-700{\rm ~cm}^{-1}$).  If some kind of general long-scale distortion of the
wavelength scale was present, we might expect to find a change of sign in
the average $\da$ when comparing, for example, the negative-shifters or
positive-shifters against the anchors (see equation \ref{eq:omega}).

From these systems we remove all transitions with $q_1$ coefficients of
`mediocre' magnitude ($300 \le \left|q_1\right| < 700{\rm ~cm}^{-1}$) so as
to clearly delineate the three types of transition (anchor, positive and
negative-shifter). We then calculate $\da$ for this sample. We then remove
each type of transition separately and calculate $\da$ again. There is a
small complication due to the presence of the occasional blend between
Cr{\sc \,ii} and Zn{\sc \,ii} $\lambda$2062 since these lines shift in
opposite directions. In cases where these two lines were blended together,
we removed both of them from all fits since they are of different `type'
and cannot be removed individually.

We present our results in Table \ref{tab:apns}. Again, precise,
quantitative comparison of the values of $\da$ obtained with the various
types of transitions removed is difficult since neither the values or the
1$\sigma$ error are independent. Also, when we remove the positive
shifters, the uncertainly in $\da$ is quite large. However, we do not find
any evidence to suggest that $\da$ is somehow dependent on the sign of
$q_1$ from this test. That is, the \hiz~results seem to be robust against
the presence of simple systematic errors. Indeed, this is corroborated by
the results of removing atmospheric dispersion effects (Table \ref{tab:da}
and Fig. 7).

\begin{table}
\caption{Results of a general test for systematic errors.  We removed the
three different {\it types} of transition from a sub-sample of the 21
\hiz~systems.  The new $\da$ is given for each case. These should be
compared to the value obtained when only transitions with mediocre $q_1$
coefficients were removed. Note that neither the values or the 1$\sigma$
errors are independent of each other.}
\begin{center}
\label{tab:apns}
\begin{tabular}{lc}\hline
Sub-sample removed & $\left<\da\right>_{\rm w}/10^{-5}$\\\hline
None              & $-0.90 \pm 0.34$\\
Mediocre-shifters & $-1.29 \pm 0.39$\\
Anchors           & $-1.46 \pm 0.44$\\
Negative-shifters & $-1.34 \pm 0.65$\\
Positive-shifters & $-1.39 \pm 1.01$\\\hline
\end{tabular}
\end{center}                            
\end{table}

\section{Conclusions}

The most important conclusion of the work described here is that after a
detailed search for instrumental and astrophysical effects, we have found no
systematic effect which can easily mimic a negative $\da$.

Of those effects we have identified and explored, two systematic effects
have been found to be {\it potentially} significant: atmospheric dispersion
and isotopic evolution.  If we were to remove these effects from the data,
the resulting $\da$ would actually be {\it more negative} than the
uncorrected effect (see Figs. 7 and 8). These corrections are {\it not}
applied to the data in M01a and W01 because (i) our estimates provide {\it
upper limits} on the magnitude of the effects, and (ii) we prefer to be as
conservative as possible as to the significance of the deviation of $\da$
from zero.

Tighter constraints on $\alpha$ are possible for realistic SNR from current
telescopes/spectrographs. The analyses above indicate that another order of
magnitude precision can be gained using our methods before wavelength
calibration and laboratory wavelength errors become significant. We hope
that M01a, W01 and the present work encourage further well calibrated,
higher SNR observations of QSOs and that an independent check on our
results can be made.

\section*{Acknowledgments}
We are very grateful to Tom Bida and Steven Vogt who provided much detailed
information about Keck/HIRES. We thank Ulf Griesmann, Sveneric Johansson,
Rainer Kling, Richard Learner, Ulf Litz\'{e}n, Juliet Pickering and Anne
Thorne for much detailed information about their laboratory wavelength
measurements. We also acknowledge helpful discussions with Michael Bessel,
Bob Carswell, Roberto De Propris, Alberto Fern\'{a}ndez-Soto, John
Hearnshaw, Alexander Ivanchik, Jochen Liske and Geoff Marcy. We are
grateful to the John Templeton Foundation for supporting this work. MTM and
JKW are grateful for hospitality at the IoA Cambridge, where some of this
work was carried out.

\label{lastpage}

\begin{thebibliography}{99}
\bibitem{bachall1}  Bahcall J. N., 1967, ApJ, 149, L7
\bibitem{bachall2}  Bahcall J. N., Sargent W. L. W., Schmidt M., 1967, ApJ,
149, L11
\bibitem{barrow}  Barrow J. D., 1987, Phys. Rev. D, 35, 1805
\bibitem{bechtold}  Bechtold J., Weymann R. J., Lin Z., Malkan M. A., 1987,
ApJ, 315, 180
\bibitem{bergeron}  Bergeron J., Stasi\'{n}ska G., 1986, A\&A, 169, 1
\bibitem{bergeron2}  Bergeron J. et al., 1994, ApJ, 436, L33
\bibitem{churchill}  Churchill C. W., 1997, Ph.D. thesis, UC Santa Cruz
\bibitem{churchill99} Churchill C. W., Rigby J. R., Charlton J. C.,
Vogt S. S., 1999, ApJS, 120, 51
\bibitem{cowie}  Cowie L. L., Songaila A., 1995, ApJ, 453, 596
\bibitem{dirac}  Dirac P. A. M., 1937, Nat, 139, 323
\bibitem{drullinger}  Drullinger R. E., Wineland D. J., Bergquist J. C.,
1980, Appl. Phys., 22, 365
\bibitem{dzubaa}  Dzuba V. A., Flambaum V. V., Webb J. K., 1999a,
Phys. Rev. Lett., 82, 888
\bibitem{dzubab}  Dzuba V. A., Flambaum V. V., Webb J. K., 1999b,
Phys. Rev. A., 59, 230
\bibitem{dzubac}  Dzuba V. A., Flambaum V. V., Murphy M. T., Webb
J. K., 2001, Phys. Rev. A, 63, 042509
\bibitem{edlen53}  Edl\'{e}n B., 1953, J. Opt. Soc. Am., 43, 339
\bibitem{edlen66}  Edl\'{e}n B., 1966, Metrologia, 2, 71
\bibitem{feretti}  Feretti L., Dallacasa D., Govoni F., Giovannini G.,
Taylor G. B., Klein U., 1999, A\&A, 344, 472
\bibitem{ferland}  Ferland G. J., 1993, University of Kentucky Department
of Physics and Astronomy Internal Report
\bibitem{filippenko}  Filippenko A. V., 1982, PASP, 94, 715
\bibitem{forgacs}  Forg\'{a}cs P., Horv\'{a}th Z., 1979, General Relativity
and Gravitation, 10, 931
\bibitem{gay}  Gay P. L., Lambert D. L., 2000, ApJ, 533, 260
\bibitem{griesmann} Griesmann U., Kling R., 2000, ApJ, 536, L113
\bibitem{hallistadius} Hallstadius L., 1979, Z. Phys. A, 291, 1220
\bibitem{ivanchik}  Ivanchik A. V., Potekhin A. Y., Varshalovich D. A.,
1999, A\&A, 343, 439
\bibitem{kupka}  Kupka F., Piskunov N. E., Ryabchikova T. A., Stempels
H. C., Weiss W. W., 1999, A\&AS, 138, 119
\bibitem{li}  Li L. -X., Gott J. R. III, 1998, Phys. Rev. D, 58, 103513
\bibitem{lopez}  Lopez S., Reimers D., Rauch M., Sargent W. L. W., Smette
A., 1999, ApJ, 513, 598
\bibitem{marciano}  Marciano W. J., 1984, Phys. Rev. Lett., 52, 489
\bibitem{milne}  Milne E. A., 1935, Relativity, Gravitation and World
Structure, Clarendon press, Oxford
\bibitem{milne2}  Milne E. A., 1937, Proc. Roy. Soc. A, 158, 324
\bibitem{moore}  Moore C. E., 1971, Atomic Energy Levels As Derived From
Analyses of Optical Spectra, Natl. Stand. Rel. Data Ser.,
Natl. Bur. Stand. (U.S.A.), Vol 1 and 2, U.S. Govt. Print Office,
Washington, D.C., U.S.A.
\bibitem{morris} Morris S. L. et al., 1986, ApJ, 310, 40
\bibitem{murphy} Murphy M. T., Webb J. K., Flambaum V. V., Dzuba V. A.,
Churchill C. W., Prochaska J. X., Barrow J. D., Wolfe A. M., 2001a, MNRAS,
accepted (M01a) (astro-ph/0012419)
\bibitem{murphy2}  Murphy M. T., Webb J. K., Flambaum V. V., Prochaska
J. X., Wolfe A. M., 2001c, MNRAS, accepted (astro-ph/0012421)
\bibitem{nave}  Nave G., Learner R. C. M., Thorne A. P., Harris C. J.,
1991, J. Opt. Soc. Am. B, 8, 2028
\bibitem{norlen}  Norl\'{e}n G., 1973, Phys. Src., 8, 249
\bibitem{palmer} Palmer B. A., Engleman R. Jr., 1983, Atlas of the Thorium
Spectrum, Los Alamos National Laboratory, Los Alamos, New Mexico
\bibitem{peck}  Peck E. R., Reeder K., 1972, J. Opt. Soc. Am., 62, 958
\bibitem{pickering}  Pickering J. C., Thorne A. P., Webb J. K., 1998,
MNRAS, 300, 131
\bibitem{pickering2} Pickering J. C., Thorne A. P., Murray J. E.,
Litz\'{e}n U., Johansson S., Zilio V., Webb J. K., 2000, 319, 163
\bibitem{piskunov}  Piskunov N. E., Kupka F., Ryabchikova T. A., Weiss
W. W., Jeffery C. S., 1995, A\&AS, 112, 525
\bibitem{prochaska1} Prochaska J. X., 1999, ApJ, 511, L71
\bibitem{prochaska3} Prochaska J. X., Wolfe A. M., 2000, ApJ, 533, L5
\bibitem{rosman}  Rosman K. J. R., Taylor P. D. P., 1998, Pure and Applied
Chemistry, 70, 217
\bibitem{songaila}  Songaila A., Cowie L. L., Hogan C. J., Rugers M., 1994,
Nat, 368, 599
\bibitem{steidel}  Steidel C. C., Sargent W. L. W., 1989, ApJ, 343, L33
\bibitem{timmes} Timmes F. X., Clayton D. D., 1996, ApJ, 472, 723
\bibitem{timmes2}  Timmes F. X., Woosley S. E., Weaver T. A., 1995, ApJS,
98, 617
\bibitem{valenti}  Valenti J. A., Butler R. P., Marcy G. W., 1995, PASP,
107, 966
\bibitem{varshalovich1}  Varshalovich D. A., Potekhin A. Y., 1995, Space
Sci. Rev., 74, 259
\bibitem{varshalovich2} Varshalovich D. A., Potekhin A. Y., Ivanchik A. V.,
2000, in Dunford R. W., Gemmel D. S., Kanter E. P., Kraessig B., Southworth
S. H., Young L., eds., AIP Conf. Proc. 506, X-ray and Inner-Shell
Processes. Argonne National Laboratory, Argonne, Illinois, p. 503
\bibitem{weast}  Weast R. C., 1979, Handbook of Chemistry and Physics (60th
edition), CRC Press, Florida, p. E-387
\bibitem{webb}  Webb J. K., Flambaum V. V., Churchill C. W., Drinkwater
M. J., Barrow J. D., 1999, Phys. Rev. Lett., 82, 884 (W99)
\bibitem{webb2} Webb J. K., Murphy M. T., Flambaum V. V., Dzuba V. A.,
Barrow J. D., Churchill C. W., Prochaska J. X., Wolfe A. M., 2001,
Phys. Rev. Lett., accepted (W01) (astro-ph/0012539)
\bibitem{whaling} Whaling W., Anderson W. H. C., Carle M. T., Brault J. W.,
Zarem H. A., 1995, J. Quant. Spectrosc. Radiat. Transfer, 53, 1
\end{thebibliography}
\end{document}